\documentclass[useAMS,usenatbib]{mn2e}
   \usepackage{graphicx}
   \usepackage{color}
   \usepackage{epstopdf}
   \usepackage{pdfsync}
\font\fiverm=cmr5             \font\sevenrm=cmr7

          \font\sixrm=cmr6       

\def\dover#1#2{\hbox{${{\displaystyle#1 \vphantom{(} }\over{
   \displaystyle #2 \vphantom{(} }}$}}
{\catcode`\@=11                                                  
\gdef\SchlangeUnter#1#2{\lower2pt\vbox{\baselineskip 0pt\lineskip0pt    
\ialign{$\m@th#1\hfil##\hfil$\crcr#2\crcr\sim\crcr}}}}           
\def\gtrsim{\mathrel{\mathpalette\SchlangeUnter>}}               
\def\lesssim{\mathrel{\mathpalette\SchlangeUnter<}}    

\def\fsc{\alpha_{\hbox{\sevenrm f}}}                                
\def\sigt{\sigma_{\hbox{\fiverm T}}}                                

\def\omegaB{\omega_{\hbox{\sixrm B}}}

\def\LB{L_{\hbox{\sixrm B}}}
\def\epsilonB{\varepsilon_{\hbox{\sixrm B}}}
\def\UB{{\cal U}_{\hbox{\sixrm B}}}
\def\ESSC{E_{\hbox{\sixrm SSC}}}

\def\teq#1{$\, #1\,$}                         % text equation
\def\pmax{p_{\mathrm{max}}} 
\def\gammax{\gamma_{\mathrm{max}}} 
\def\ThetaBfone{\Theta_{\hbox{\sixrm Bf1}}}
\def\betaoneHT{\beta_{\hbox{\sixrm 1HT}}}

\title[Probing Acceleration and Turbulence at Relativistic Shocks in Blazar Jets]
{Probing Acceleration and Turbulence at Relativistic Shocks in Blazar Jets}
\author[M.~G. Baring, M. B\"ottcher \& E. J. Summerlin]{
Matthew G. Baring,$^{1}$\thanks{E-mail:
baring@rice.edu (MGB); Markus.Bottcher@nwu.ac.za (MB); errol.summerlin@nasa.gov (EJS)}
Markus B\"ottcher$^{2}$ and Errol J. Summerlin$^{3}$\\
$^{1}$Department of Physics and Astronomy - MS 108, Rice University,
6100 Main Street, Houston, Texas 77251-1892, USA\\
$^{2}$Centre for Space Research, North-West University,
Potchefstroom Campus, Potchefstroom, 2520, South Africa\\
$^{3}$Heliospheric Physics Laboratory, Code 672,
NASA's Goddard Space Flight Center, Greenbelt, MD 20770, USA}

\begin{document}

\date{Accepted 2016 September 12, Received 2016 September 12; in original form 2016 June 10}

\pagerange{\pageref{firstpage}--\pageref{lastpage}} \pubyear{2016}

\maketitle

\label{firstpage}

\begin{abstract}
Diffusive shock acceleration (DSA) at relativistic shocks is widely
thought to be an important acceleration mechanism in various
astrophysical jet sources, including radio-loud active galactic nuclei
such as blazars.  Such acceleration can produce the non-thermal
particles that emit the broadband continuum radiation that is detected
from extragalactic jets.  An important recent development for blazar
science is the ability of {\it Fermi}-LAT spectroscopy to pin down
the shape of the distribution of the underlying non-thermal
particle population.  This paper highlights how multi-wavelength spectra
spanning optical to X-ray to gamma-ray bands can be used to probe
diffusive acceleration in relativistic, oblique, magnetohydrodynamic
(MHD) shocks in blazar jets.  Diagnostics on the MHD turbulence near
such shocks are obtained using thermal and non-thermal particle
distributions resulting from detailed Monte Carlo simulations of
DSA.  These probes are afforded by the characteristic property that the
synchrotron $\nu F_{\nu}$ peak energy does not appear in the gamma-ray
band above 100 MeV.   We investigate self-consistently the radiative
synchrotron and inverse Compton signatures of the simulated particle
distributions.  Important constraints on the diffusive mean free paths of
electrons, and the level of electromagnetic field turbulence are
identified for three different case study blazars, Mrk 501, BL Lacertae
and AO 0235+164.  The X-ray excess of AO 0235+164 in a flare state can
be modelled as the signature of bulk Compton scattering of external
radiation fields, thereby tightly constraining the energy-dependence of
the diffusion coefficient for electrons.  The concomitant interpretations
that turbulence levels decline with remoteness from jet shocks, and the
probable significant role for non-gyroresonant diffusion, are posited.
\end{abstract}


\section{Introduction}

Extragalactic jets of collimated relativistic outflows are some of the
most powerful emitters of radiation in the Universe.  Two contrasting
types of jets are found in transient gamma-ray bursts (GRBs), and
persistent but highly-variable active galactic nuclei (AGN).  Both are
considered prime candidates for the production of ultra-high energy
cosmic rays above \teq{10^{18}}eV, and perhaps also the high energy
neutrinos recently detected by IceCube \citep{Aartsen14}. Yet they are
very different in their origin. GRBs probably result from the explosion
of massive progenitor stars in {\it hypernovae}, or the merger of
compact binary neutron stars, whereas AGN are continually powered over
eons by material accreted onto supermassive black holes (SMBHs) in the
center of distant galaxies.  The masses of such SMBHs are typically in
the range \teq{10^{6.5}-10^9 M_{\odot}}
\citep{Bentz-Katz15},\footnote{For an accessible Web database of SMBH
mass listings, see {\tt http://www.astro.gsu.edu/AGNmass}.} deduced in
part from their large luminosities, typically
\teq{10^{42}-10^{47}}erg/sec, and also from reverberation mapping
techniques \citep[e.g.,][]{Peterson04}.  In this paper, the focus is on
the interpretation of the jet environments of AGN, in particular on the
specific subset known as {\it blazars}, which exhibit flares with short
timescale variations in radio, optical, X-ray and $\gamma$-ray
wavebands.  The class of blazars was identified following the discovery
\citep{Hartmann92,Lin92} of transient gamma-ray emission in 3C 279 and
Mrk 421 by the EGRET instrument on the Compton Gamma-Ray Observatory in
the 100 MeV -- 1 GeV range. This was around the same time that the
Whipple atmospheric \v{C}erenkov telescope (ACT) discerned that Mrk 421
also emitted at TeV energies \citep{Punchetal92}.

Blazars generally evince the key identifying feature of their parent BL
Lac objects like {\it BL Lacertae} of a general absence of emission
lines in their non-thermal optical spectra.  Their radio continua often
possess quite flat spectra. These characteristics suggest synchrotron
radiation from non-thermal electrons gyrating in magnetic fields as the
origin of their emission in these two wavebands.  Such a hypothesis is
supported by detections of polarization in both radio and optical bands.
 Most remarkable among the ensemble of blazar polarimetry studies is the
observation of rapid optical polarization angle swings in 3C 279
\citep[seen also for other blazars, including PKS
1510-089;][]{Marscher10}, contemporaneous with a gamma-ray flare seen by
{\it Fermi}-LAT \citep{Abdo10b}.  This indicates a large-scale ordering
of the magnetic field in the 3C 279 jet, and spatial coincidence of the
optical and gamma-ray emission zones. In many blazars, this putative
synchrotron component extends to X-rays, where polarization measurements
in the not-to-distant future define a hope for further constraining the
emission mechanism of blazars \citep[e.g.,][]{Krawczynski11}.

The prevailing paradigm is that blazars' gamma-ray signals are generated
by inverse Compton scattering of synchrotron photons by the same
relativistic electrons that emit this radio-to-X-ray signal, so-called
synchrotron-self-Compton (SSC) models \citep[e.g.,][]{MCG92,MK97,CB02}.
An alternative possibillity for the target photons seeding this
upscattering is an IR/optical/UV photon source of possibly disk or
ambient origin \citep[e.g.,][]{DSM92,SBR94} not too distant from the
black hole. The main competing scenario for $\gamma$-ray production in
blazars is the hadronic model \citep[e.g.,][]{MB89,RM98,MP01} where
non-thermal protons collide with disk and jet-associated synchrotron
photons, generating pair cascades via photo-pion/muon production
processes. Proton synchrotron radiation can also appear in the
$\gamma$-ray band. High energy neutrinos are a signature product of such
cascades, forging a connection of hadronic blazar models to the
observation of energetic neutrinos up to around \teq{10^{15}}eV by the
IceCube experiment \citep{Aartsen14}.

Blazar spectra observed above 300 GeV by ACTs are typically quite steep,
defining a turnover.  These photons are subject to strong
\teq{\gamma\gamma} pair absorption in propagating to Earth due to the
presence of intergalactic infra-red and optical starlight background
fields.  Correcting for such attenuation can yield inferences of very
flat intrinsic source spectra \citep[e.g.,][]{SBS07}, yet uncertainties
in the background fields permeate such protocols. An important recent
development for blazar science has been the improvement of sensivity in
the 100 MeV-100 GeV window, afforded by the {\it Fermi}-Large Area
Telescope (LAT).  Over the last eight years, LAT data has enabled
measurements of the power-law index of spectra from numerous blazars
\citep{Abdo09,Abdo10c}.  This provides a more robust measure of the
underlying non-thermal particle population, since the LAT band is
generally subject to only small \teq{\gamma\gamma} pair attenuation, and
negligibly so below around 3 GeV. This is a new probe enabled for
multiwavelength blazar studies, a tool that we exploit to some extent
here. Yet more particularly, one of our case studies, AO 0235+164, is
not yet detected with current ACT facilities, so that the {\it
Fermi}-LAT data prove critical to our interpretation of its jet
environment.

The rapid flux variability seen in radio, X-ray and GeV--TeV flares
\citep[e.g., see][for Mrk 421]{MFTetal99,TKMetal00} drives the
prevailing picture for the blazar environment: their jets are
relativistic, and compactly structured on small spatial scales that are
unresolvable by present gamma-ray telescopes. For a survey of radio and
X-ray jet properties including spatial morphology, see \citet{HK06}. 
The supersonic outflows in these jets naturally generate relativistic
shocks, and these can form the principal sites for dissipation of the
ballistic kinetic energy via acceleration of electrons and perhaps also
ions to the ultrarelativistic energies demanded by the X-ray and
$\gamma$-ray data.  Diffusive Fermi acceleration at such jet shocks is
believed to be the main candidate mechanism for energizing such charges
\citep{BE87,Drury83,JE91}. This is motivated by the fact that Fermi
acceleration is both extremely efficient and very fast, precipitating
acceleration rates \teq{{\dot \gamma}} of the order of the gyrofrequency
\teq{\omega_g = eB/mc}.  This follows because the diffusive collisions
of electrons and ions with MHD turbulence in jets is putatively
dominated by gyroresonant interactions.  The efficiency of
particle-turbulence interactions is an issue that we will visit in this
paper.  We note that shocks also may span a large surface cross section
of the jet that the outward-propagating plasma may encounter with high
probability.

Another possibility is that shear layers encapsulating sharp velocity
gradients transverse to the net flow may precipitate Fermi-type
acceleration \citep{Ostrowski90} due to transport of charges straddling
the shear ``discontinuity.'' Observational evidence supporting such
transverse velocity structure comes from parsec-scale limb-brightening
of blazar and radio galaxy jets revealed in VLBI observations
\citep[e.g., see ][for Mrk 501]{GGFetal04}. MHD/hydrodynamic simulations
of spine-sheath jets and their launching
\citep[e.g.,][]{MK07,MN07,MHN07,TMN08} indicate that the sheath, in
combination with a poloidal magnetic field, aids in stabilizing the jet.
 Rayleigh-Taylor-type instabilities may develop at the spine-sheath
interface, and such turbulence offers another promising avenue for
accelerating relativistic particles. Indeed, particle-in-cell
\teq{e^{\pm}} plasma simulations of relativistic shear flows
\citep{LBS13} indicate substantial energization of pairs in the boundary
layers. Other paradigms such as acceleration by reconnection of magnetic
fields embedded in Poynting-flux dominated outflows can be envisaged. 
Whether shocks or shear boundaries or reconnection zones provide the
dominant energization site for non-thermal particles in blazars defines
a major objective for future theoretical efforts within the blazar
community.

Our considerations here will focus on acceleration at blazar jet shocks.
To explore the blazar shock paradigm, the standard practice has been to
develop multiwavelength (MW) spectral models, spanning radio to TeV
gamma-ray wavelengths \citep{DSM92,SBR94,MCG92,MK97,CB02,BDF08}. This
generally gives a broad-brush assessment of a range of jet environment
parameters, but the spectral fits in any one band are of limited
quality. The high statistics spectroscopy afforded by the {\it
Fermi-LAT} data demands a closer look at the constraints observations
can elucidate for the shocked environs of blazar jets and on the shock
acceleration process itself. In particular, theoretical studies of shock
acceleration have determined \citep[e.g.,][]{EJR90,KH89,ED04,SB12} that
a wide variety of power-law indices for accelerated charges are possible
for a given velocity compression ratio across the discontinuity.  These
indices are sensitive to the orientation of the mean magnetic field, and
the character of the {\it in situ} MHD turbulence.  In addition, it has
been understood for nearly two decades \citep[see][]{IT96} that
diffusive shock acceleration is so efficient that low levels of field
turbulence are required in blazar jets to accommodate synchrotron
spectral peaks appearing in the X-ray band. This serves as a central
issue to the discourse of this paper.

In the multi-wavelength blazar models developed herein, detailed results
from comprehensive Monte Carlo simulations of diffusive acceleration at
relativistic shocks are employed, building upon the exposition of
\citet{SB12}. These simulations capture the relationship between
turbulence parameters and the power-law index, and also the connection
between the thermal bulk of the population (a hot Maxwellian-like
component) and the power-law tail of the accelerated species. Such an
approach goes beyond an elemental description of the distribution of
accelerated charges, usually a power-law truncated at some minimum and
maximum particle Lorentz factors, that is commonly invoked in blazar
spectral modeling. Our shock simulation approach is outlined in
Section~\ref{sec:shock_accel}. We fold these simulation results through
the one-zone SSC/external Compton models of \citet{Boettcher13} for
radiation emission and transport in blazars. The radiation modeling
protocols are summarized in Section~\ref{sec:radmodels}, and
subsequently results from this analysis are presented therein.

Diagnostics are obtained for the particle mean free path and the level
of field turbulence for our three chosen blazars, namely Mrk 501, BL
Lacertae, and the BL Lac object AO 0235+164. The non-thermal emission
components {\it are primarily generated by leptons that undergo repeated
drift acceleration and interspersed upstream reflections in the shock
layer}, the result being extremely flat distributions emerging from the
acceleration zones. By exploring a range of dependences of the diffusive
mean free path \teq{\lambda} on the momentum \teq{p} of accelerated
charges, the possible interpretation that turbulence levels decline with
remoteness from a shock is identified, perhaps signalling a significant
role for non-gyroresonant diffusion in the vicinity of blazar jet
shocks, an interpretation discussed in Section~\ref{sec:discussion}. 
This determination is impelled by the fact that gyroresonant diffusive
acceleration and energization in magnetic reconnection zones are just
too fast to accommodate the multi-wavelength spectra of blazars, in
particular the positioning of the synchrotron \teq{\nu F_{\nu}} peak in
the optical or X-ray bands.

\newpage

\section[]{Simulations of Diffusive and Drift Acceleration at Shocks}
 \label{sec:shock_accel}

To construct a more precise assessment of multiwavelength emission 
models for blazars, and derive a deeper understanding of their jet physics, 
it is necessary to go beyond simplistic truncated power-law distributions 
for the radiating leptons and hadrons.  This is our approach in this paper, 
where detailed modeling of particle distribution functions, anisotropies, 
and their relationship to plasma turbulence are afforded through 
simulations of shock acceleration processes.  Simulations are 
the most encompassing technique to apply to the blazar jet problem 
because the shocks are inherently relativistic: faster and slower regions 
of the jet flow travelling near \teq{c} have relative velocities that are 
mildly-relativistic.  This domain will form the focus of our exposition here. 

\subsection[]{Background on Shock Acceleration Theory}
 \label{sec:shock_theory}

Various approaches have been adopted to model diffusive shock
acceleration (DSA) at relativistic discontinuities. These include
analytic methods \citep[e.g.,][]{Peacock81,KS87,KH89,Kirk00}, and Monte
Carlo simulations of convection and diffusion
\citep[e.g.,][]{EJR90,BO98,ED04,NO04,SB12}. Particle-in-cell (PIC)
plasma simulations have also been employed to study this problem via
modeling the establishment of electromagnetic turbulence
self-consistently
\citep[e.g.,][]{Gallant92,Silva03,Nishikawa05,Spitk08,SS09} in
conjunction with the driver charges and their currents.  Important
insights are gleaned from each of these techniques. All exhibit the core
property that in collisionless shocks, non-thermal, charged particles
gain energy by scattering between MHD turbulence (for example magnetic
``islands'' observed in PIC simulations) that is ``anchored'' in the
converging upstream and downstream plasmas, the so-called {\it Fermi
mechanism}. This defines a fundamental association between turbulence
contained in shock environs, diffusion and ultimate acceleration.

Each of these approaches to the shock acceleration problem has both
merits and limitations.  Analytic techniques provide core insights into
global character, though are often restricted to treating particles well
above thermal energies where the acceleration process possesses no
momentum scale.  Monte Carlo simulations with prescribed injected
turbulence \citep[e.g.,][]{NO04} explore wave-particle interactions and
diffusive acceleration well above thermal energies, but do not treat the
electrodynamics of wave generation and dissipation self-consistently.  
Particle-in-cell simulations provide the greatest depth in modeling
shock layer microphysics by solving Maxwell's equations and the
Newton-Lorentz force equation, though their macro-particle approximation
does eliminate electrodynamic information on the smallest scales.  With
increased box sizes, PIC codes have now realized the establishment of
truly non-thermal components, but still cannot explore acceleration
beyond modest non-thermal energies \citep[e.g.,][]{SS09}; their focus is
still on the thermal dissipation and injection domains in shock layers. 
The Monte Carlo approach employed here subsumes the microphysics in a
parametric description of diffusion, and so does not investigate the
feedback between charges, currents and hydromagnetic waves. Yet it is
ideal for describing the diffusive and shock drift acceleration of
particles from thermal injection scales out to the highest energies
relevant to blazar jet models, and so is the preferred tool for
interfacing with astronomical emission datasets.

A key feature of both relativistic and non-relativistic shock
acceleration theory is that the acceleration process possesses no
momentum scale, and the resulting particle distribution takes the form
\teq{dn/dp \propto p^{-\sigma}}.  For non-relativistic shocks, since
their speeds \teq{v\approx c} far exceed \teq{u_{1x}} (\teq{u_{2x}}),
the upstream (downstream) flow speed component in the co-ordinate
direction \teq{x} normal to the shock, the energetic particles are
nearly isotropic in all fluid frames.  The acceleration process then
establishes \citep[e.g.,][]{Drury83,JE91} a power-law distribution with
index \teq{\sigma = (r+2)/(r-1)}, where \teq{r=u_{1x}/u_{2x}} is the
shock's velocity compression ratio. The index \teq{\sigma} in this
\teq{u_1\ll c} limit is independent of the shock speed, \teq{u_1}, the
upstream field obliquity angle \teq{\ThetaBfone} (to the shock normal,
which is in the \teq{x}-direction throughout this paper), and any
details of the scattering process. This canonical result has propelled
the popularity of shock acceleration as a key element of paradigms
promoted over the last four decades for the generation of cosmic rays.

In contrast, it is widely understood that because plasma anisotropy is
prevalent in relativistic shocks, when \teq{u_1\sim c}, the index
\teq{\sigma} of the power-law distribution is a function of the flow
speed \teq{u_1}, the field obliquity angle \teq{\ThetaBfone}, and the
nature of the scattering. Test-particle acceleration in parallel
(\teq{\ThetaBfone =0}) relativistic shocks evinces the essential
property that a so-called ``universal'' spectral index \teq{\sigma\sim
2.23} exists in the two limits of \teq{\Gamma_1 \gg 1} and small angle
scattering, i.e., \teq{\delta \theta \ll 1/\Gamma_1}, for a shock
compression ratio of \teq{r=u_{1x}/u_{2x}=3}.  This was showcased in the
seminal work \citet{Kirk00}, which employed semi-analytic methods to
solve the diffusion-convection equation, and was also generated in the
Monte Carlo analyses of \citet{BO98} and \citet{Baring99}. Here
\teq{\delta\theta} is the average angle a particle's momentum vector
deviates in a scattering ``event,'' and $\Gamma_1 = (1 -
[u_1/c]^2)^{-1/2}$ is the Lorentz factor of the upstream flow in the
shock rest frame.  For all other parameter regimes in relativistic
shocks, a wide range of departures of \teq{\sigma} from this special
index is observed \citep[see][and references therein]{ED04,SB12}. While
this is a complication, it enables powerful spectral diagnostics on the
large scale electromagnetic structure of shocks and also the MHD
turbulence in their environs.

\subsection[]{The Monte Carlo Simulational Method}
 \label{sec:MC}

The Monte Carlo technique employed here to model acceleration at blazar
shocks solves the Boltzmann equation with a phenomenological scattering
operator \citep[][]{JE91,EJR90,SB12}.  The simulation space is divided
into grid sections with boundaries parallel to the planar shock
interface.  Sections can possess different flow velocities, mean
magnetic field vectors, etc., defining the MHD structure of the shock.
For electron-ion shocks, the vastly different inertial scales can be
easily accommodated, though the present application is for pair plasma
shocks. Charges are injected far upstream and diffuse and convect in the
shock neighborhood. Their flux-weighted contributions to the momentum
distribution function are logged at any position \teq{x}. Each particle
is followed until it leaves the system by either convecting sufficiently
far downstream, or exceeding some prescribed maximum momentum
\teq{\pmax}.  To enhance the speed of the code, a {\it probability of
return} boundary is introduced \citep[following][]{EJR90} at several
diffusion lengths downstream of the shock, beyond which the decision to
retain or discard a charge subject to convection and diffusion is made
statistically: see \citet{SB12}.

The details of particle transport are given in \citet{EJR90,ED04,SB12}:
the simulation can model both scatterings with small angular deflections
\teq{\delta\theta \ll 1/\Gamma_1} (seeding pitch angle diffusion) and
large ones (\teq{\delta\theta \gtrsim 1/\Gamma_1}) using several
parameters. These deflections can be viewed as a mathematical
discretization of charge trajectory segments in electromagnetic
turbulence: the shorter the segment, the smaller \teq{\delta\theta} is.
The scatterings are assumed to be quasi-elastic in the local fluid
frame, an idealization that is usually valid because in blazar jets and
other astrophysical systems the flow speed far exceeds the Alfv\'{e}n
speed (true for the models in Table~1), and contributions from
stochastic second-order Fermi acceleration are small.  In this paper,
the focus will be on small angle \teq{\delta\theta \ll 1/\Gamma_1}
domains, where {\it pitch angle diffusion} is realized; results for
large angle scattering are illustrated in \citet{EJR90} and
\citet{SBS07}. The simulation assumes that the complicated plasma
physics of wave-particle interactions can be described by a simple
scattering relation for particles, viz.
\begin{equation}
   \lambda_{\parallel} \; =\; \lambda_1 \, \dover{\rho_1}{\rho} \left ( \dover{r_g}{r_{g1}} \right )^\alpha
   \;\equiv\; \eta_1\, r_{g1} \left( \dover{p}{p_1} \right)^{\alpha}
   \;\; ,  \quad
   \kappa_{\parallel} = \lambda_{\parallel} v/3\ ,
 \label{eq:mfp_alpha}
\end{equation}
where \teq{v=p/m} is the particle speed in the local upstream or
downstream fluid frame, \teq{r_g =pc/(QeB)} is the gyroradius of a
particle of charge $Qe$, and $\rho$ is the plasma density, with a far
upstream value of \teq{\rho_1}. Here, \teq{\lambda_\parallel}
(\teq{\kappa_{\parallel}}) is the mean free path (spatial diffusion
coefficient) in the local fluid frame, parallel to the field {\bf B},
with \teq{\lambda_\parallel\gtrsim r_g} being a fundamental bound (Bohm
limit) for physically meaningful diffusion.  Scalings of the momentum
\teq{p_1=mu_{1x}=m\beta_{1x}c} and the mean free path \teq{\lambda_1=
\eta_1 r_{g1}} for \teq{r_{g1} =p_1c/(QeB_1)} are introduced to simplify
the algebra in Eq.~(\ref{eq:mfp_alpha}).  In the local fluid frame, the
time, \teq{\delta t_f}, between scatterings is coupled \citep{EJR90} to
the mean free path, \teq{\lambda_{\parallel}}, and the maximum
scattering (i.e. momentum deflection) angle, \teq{\delta\theta} via
\teq{ \delta t_f\approx \lambda_{\parallel} \,\delta\theta^{2}/(6v)} for
particles of speed \teq{v\approx c}.

Scattering according to Eq.~(\ref{eq:mfp_alpha}) is equivalent to a
kinetic theory description \citep[e.g.,][]{FJO74,Jokipii87} where the
diffusion coefficients perpendicular to (\teq{\kappa_{\perp}}) and
parallel to (\teq{\kappa_{\parallel}}) the local field vector {\bf B}
are related via 
\begin{equation}
   \kappa_{\perp} \; =\;  \dover{\kappa_{\parallel} }{1 + (\lambda_{\parallel}/r_g)^2 } \quad .
 \label{eq:kinetic_theory}
\end{equation}
The parameter \teq{\eta \equiv \lambda_{\parallel}/r_g \propto p^{\alpha-1}} 
then characterizes the ``strength" of the scattering and the
importance of cross-field diffusion:  when \teq{\eta \sim 1} at the Bohm
diffusion limit, \teq{\kappa_{\perp} \sim \kappa_{\parallel}} and
particles diffuse across magnetic field lines nearly as quickly as along
them. This Bohm case corresponds to extremely turbulent fields, whose
fluctuations satisfy \teq{\delta B/B\sim 1}.  Each scattering event is
an elastic deflection of the fluid frame momentum vector {\bf p} through
angle \teq{\sim \delta\theta}, so that the number of deflections
constituting \teq{\lambda_\parallel}, a large angle deflection scale
(i.e., a turnaround through angle \teq{\sim \pi /2}), is proportional to
\teq{(\delta\theta)^{-2}}. Values in the range \teq{1/2 \lesssim \alpha
\lesssim 3/2} for this diffusion index are inferred from observations of
turbulence in disparate sites in the solar wind, and also from hybrid
plasma simulations, a context that will be discussed later in
Section~\ref{sec:discussion}. We remark here that the story that will
unfold in this paper is that blazar jets may present a picture with some
similarities to the solar wind environment, with higher values of
\teq{\alpha} in some cases that are associated with the large dynamic
ranges of momenta and mean free paths in their relativistic jets.

\begin{figure}
\centerline{\includegraphics[width=9.5cm]{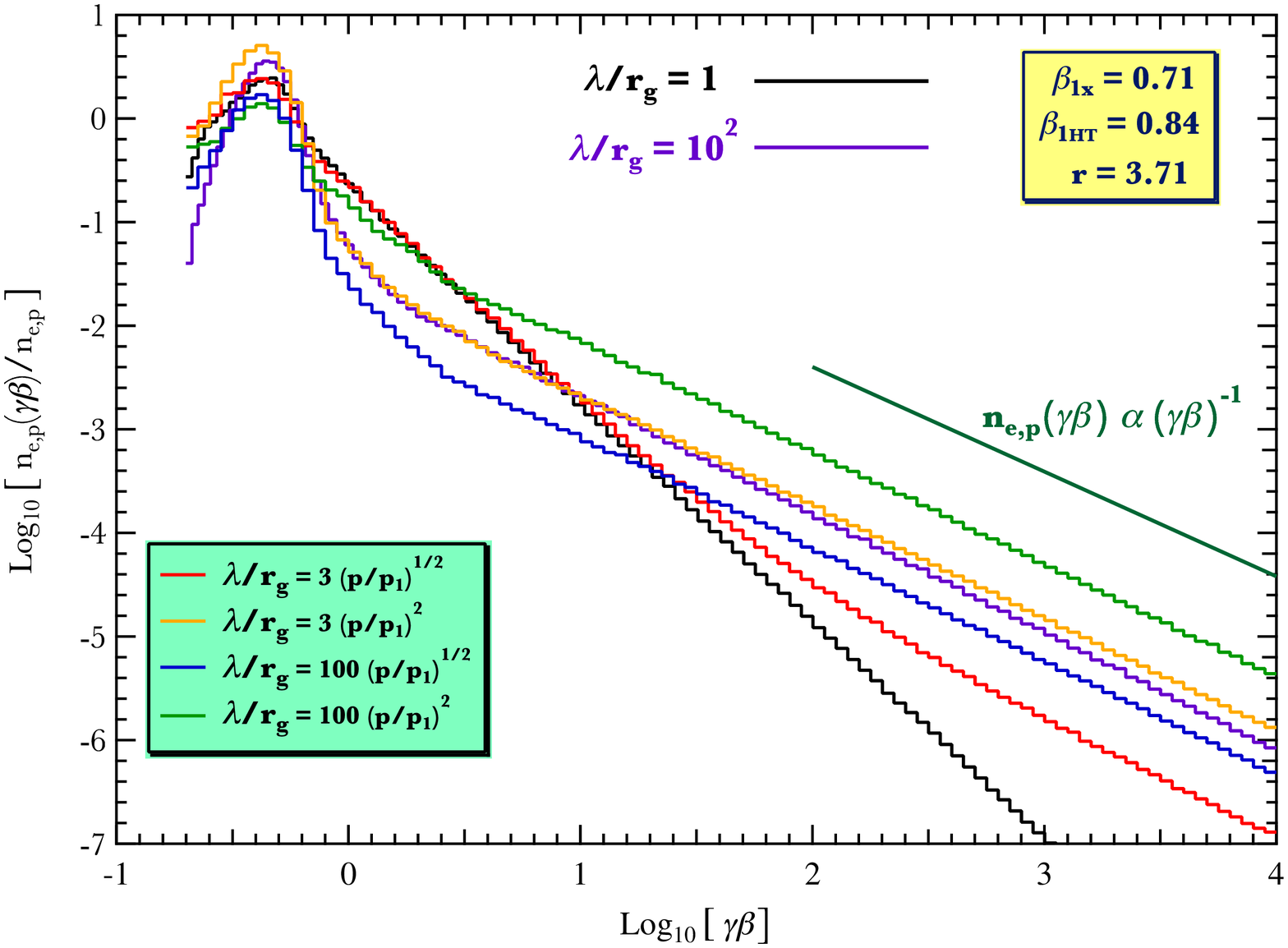}}
\vspace*{-5pt}
\caption{Particle distributions \teq{n_s(\gamma\beta ) \equiv m_s c\, dn_s/dp}
(normalized; \teq{s=e,p}) in momentum space for acceleration simulation
runs in the small angle scattering limit, corresponding to strong
mildly-relativistic shocks of upstream flow speed \teq{\beta_{1x}\equiv
u_{1x}/c =0.71}.  Here the de Hoffmann-Teller (HT) frame upstream flow
speed was set at \teq{\betaoneHT =0.84=\beta_{1x}/\cos\ThetaBfone}, with
\teq{\ThetaBfone \approx 32.3^{\circ}} being the upstream field
obliquity to the shock normal in the HT frame.  Distributions are
displayed for six different forms for the momentum dependence of the
diffusive mean free path \teq{\lambda\equiv\lambda_{\parallel}}, namely
\teq{\lambda/r_g \propto p^{\alpha -1}} with \teq{\alpha -1 = 0, 1/2}
and \teq{2}, as labelled --- see Eq.~(\ref{eq:mfp_alpha}) and associated
discussion.  The shock velocity compression ratio was fixed at
\teq{r=u_{1x}/u_{2x}=3.71}, and the upstream temperature corresponded to
a sonic Mach number of \teq{{M}_{\rm S}\sim 4}. See Fig.~10 of
\citet{SB12} for more \teq{\alpha =1} cases. At high momenta \teq{p\gg
p_1}, many of the distributions are close to the flat
\teq{n_s(\gamma\beta )\propto 1/(\gamma\beta )} asymptote that is
highlighted in dark green.
\label{fig:accel_dist}}
\end{figure}

In relativistic shocks, the distribution functions of accelerated
charges are sensitive to the choices of both the
\teq{\eta_1=\lambda_1/r_{g1}} and \teq{\alpha} diffusion parameters.
This property is illustrated in Fig.~\ref{fig:accel_dist} for the
mildly-relativistic shock domain that is germane to our blazar study.
This sensitivity is addressed at length in \citet{SB12}, where the
\teq{\alpha = 1} restriction was adopted for simplicity, so that
\teq{\lambda_{\parallel} \propto r_g} and a single value of \teq{\eta}
applies for all particle momenta. In subluminal shocks, for which
\teq{u_{1x}/\cos\ThetaBfone <c}, i.e., those where a de Hoffman-Teller
(HT) frame \citep{HT50} can be found, distributions \teq{dn/dp \propto
p^{-\sigma}} generally possess indices in the range \teq{1 < \sigma <
2.5}. The HT frame is that where the shock is at rest and the fluid
flows along {\bf B} at all positions upstream and downstream. Since
\teq{\ThetaBfone} is the upstream field obliquity angle to the shock
normal in the HT frame, and because mildly-relativistic shocks are {\it
de rigueur} for structures in blazar jets, this angle can be quite
modest: subluminal shocks are a real possibility --- see Table 1 for the
\teq{\ThetaBfone} values used in our multiwavelength spectral fitting
studies. In contrast, superluminal shocks that possess higher
obliquities have steep distributions with \teq{\sigma > 2.5} when
diffusion is not near the Bohm limit; in these circumstances, rapid
convection downstream of the shock spawns high loss rates from the
acceleration process \citep[e.g.,][]{BK90,ED04,SB12}.

\subsubsection{Representative $e^-$ Distributions for Blazar Studies}

Examples of both constant and momentum-dependent \teq{\lambda /r_g} are
depicted in Fig.~\ref{fig:accel_dist}, where the power-law tails
smoothly blend into the upper end of the thermal distribution.  These
distribution results were normalized to unit integrated density in all
cases but one, with the \teq{\lambda /r_g = 100\, (p/p_1)^{1/2}} example
being normalized to \teq{n_s=1/2} so as to aid clarity of the Figure.
For much of the fairly restricted range of obliquities corresponding to
\teq{u_1/\cos\ThetaBfone <c}, when \teq{\eta} is a constant for all
momenta, the larger the value of \teq{\eta = \eta_1}, the smaller is
\teq{\sigma}.  This flattening trend is clearly represented by the
comparison of the Bohm case (black) and the \teq{\eta =100} case
(purple) in the Figure. In fact, when \teq{\eta\gg 1}, then indices
\teq{\sigma\sim 1} are generally observed \citep{SB12} and the \teq{p\gg
p_1} distribution is extremely flat.  The origin of this behavior was
found to be {\bf shock drift acceleration} (SDA), where charges with
select gyrational phases incident upon the shock from upstream, are
trapped and repeatedly reflected back upstream by the shock
discontinuity. \citet{SB12} observed that the shock drift process is
quickly disrupted when the turbulence levels increase and \teq{\eta}
drops below around 100 or so.

Shock drift acceleration is a phenomenon that has been well-studied in
the context of non-relativistic heliospheric shocks
\citep{Jokipii82,DV86}. It has also been observed in recent PIC
simulations of non-relativistic electron-proton shocks
\citep{Park12,Park15}, albeit primarily as a source of pre-injection of
protons into the Fermi process that is manifested therein on large
scales that sample the turbulent magnetic structures. The asymmetry of
the drift electric fields straddling the discontinuity leads to a net
energization during a gyrational reflection event.  Successive episodes
of upstream excursions and reflection at the shock precipitate efficient
acceleration and postpone convective loss downstream for many shock
interaction cycles. This extensive confinement to the shock layer
automatically generates flat distributions with \teq{\sigma\sim 1-1.5}
\citep{SB12}, in a sense analogous to the blazar jet shear layer studies
of \citet{Ostrowski90}, and more or less commensurate with distribution
indices realized in some magnetic reconnection models
\citep[e.g.][]{Cerutti12}.   Only when the field obliquity
\teq{\ThetaBfone} is high enough that the shock is superluminal, and
\teq{u_{1x}/\cos\ThetaBfone >c}, does the powerful convective action of
the flow overwhelm reflection, and the acceleration process effectively
shuts off \citep{SB12,BK90,ED04}, with the index \teq{\sigma} increasing
rapidly above \teq{3-4}.  Such distributions are nominally too steep to
accommodate spectra from {\it Fermi}-LAT blazar data \citep[e.g.,
see][]{Abdo10a}, and so it is concluded that blazar jets must possess
subluminal or ``marginally luminal'' relativistic shocks.  This is the
MHD shock phase space that is explored exclusively in this paper.

These acceleration distribution results are extended here to cases where
\teq{\eta (p)} is an increasing function of momentum \teq{p}, as in
Eq.~(\ref{eq:mfp_alpha}), with representative \teq{\alpha >1} examples
being displayed in Fig.~\ref{fig:accel_dist}.  The subluminal shock
parameters used therein are mostly those in Fig.~10, left panel, of
\citet{SB12}.  The distributions all display the generic trait of a
dominant thermal population with a power-law tail that extends as high
as the geometric scale of the diffusive acceleration zone permits: this
environmental parameter is addressed in Section~\ref{sec:radmodels}. 
Each distribution possesses an injection efficiency from thermal into
the Fermi process, i.e. for slightly supra-thermal momenta \teq{p\sim
0.3p_1-3p_1}, that reflects a combination of the values of \teq{\eta}
and \teq{\alpha} in Eq.~(\ref{eq:mfp_alpha}).  The ultimate power-law
index at high momenta \teq{p\gg p_1} is determined by large values for
\teq{\eta (p)} in all these simulation runs, i.e., realizing
\teq{\sigma\sim 1} domains due to efficient shock drift acceleration in
putatively weak turbulence.

The key new spectral signature presented here is the appearance of
``flattening'' breaks in the high energy tail that are manifested when
\teq{\eta (p)} begins to exceed values around \teq{10-30}; see the
\teq{\eta_1=3}, \teq{\alpha =3/2} and \teq{\eta_1=3}, \teq{\alpha =2}
cases in Fig.~\ref{fig:accel_dist}. Hence, \teq{\alpha >1} models can
possess relatively inefficient injection at thermal energies, with
dominant thermal components, but exhibit efficient acceleration at
momenta \teq{p\gtrsim 100p_1}. The properties of plasma turbulence that
might generate \teq{\alpha >1} circumstances are discussed below in
Section~\ref{sec:discussion}, where the blazar spectral modeling that is
presented in Section~\ref{sec:radmodels} is interpreted.

\section[]{Multiwavelength Radiation Emission Modeling}
 \label{sec:radmodels}

In this section, the focus is on modeling the broadband continuum
emission of the blazars Mrk 501, BL Lacertae and  AO 0235+164 using
specific thermal plus non-thermal electron/pair distributions generated
by the Monte Carlo technique.  There are quite a few shock parameters
that can be varied, so to maximize our insights, we keep the MHD
Rankine-Hugoniot structure of the shock identical for all runs, and vary
mostly the magnitude and direction of the magnetic field, and the
diffusion parameters \teq{\eta_1} and \teq{\alpha} highlighted in
Eq.~(\ref{eq:mfp_alpha}).  Accordingly, flow speeds, compression ratios
and upstream plasma temperatures are fixed so as to reduce the myriad of
model possibilities.

The objective is to assess how the global character of the radio to
X-ray to gamma-ray data can provide insights into the nature of the
accelerator and its plasma environment.  To facilitate this goal, we
concentrate solely on leptonic models
\citep{MCG92,MK97,CB02,Boettcher13}, those with synchrotron emission
predominantly at frequencies below \teq{10^{19}}Hz, and with inverse
Compton emission dominating the gamma-ray signal. Hadronic models are
also of great interest, but then introduce more parameters, and so don't
tend to provide constraints that are as restrictive as those explored
here.  We remark that the critical synchrotron constraints on the values
of \teq{\eta_1} and \teq{\alpha} will apply to either leptonic or
hadronic scenarios for the gamma-ray signals.

\subsection{Radiation Code and Geometry Essentials}

To evaluate the light emission from the complete thermal+non-thermal
electron distributions generated by the Monte Carlo simulation described
in the previous subsection, we adopt radiation modules from the one zone
blazar radiation transfer code of \citet{Boettcher13} and \citet{BC02}.
This code treats synchrotron emission, synchrotron self-absorption at
low radio frequencies, and also inverse Compton scattering of both
synchrotron and external (nominally of disk/torus origin) seed photons.
Bremsstrahlung is modeled in the code, but is usually insignificant in
blazars: it is so for all our case study blazars, Mrk 501, BL Lac and AO
0235+164. The emission components are computed for isotropic
distributions in the jet frame and then blue-shifted and Doppler-boosted
to the observer frame using the bulk Lorentz factor \teq{\Gamma_{\rm
jet}} of the jet, and the observing angle \teq{\theta_{\rm obs}}
relative to its axis.

Note that the distributions generated in the shock acceleration
simulations are not isotropic in all reference frames. The dominant,
compressed field is located downstream of a shock, and this is the zone
where one anticipates the major contribution to both synchrotron and SSC
emission originates. Since the diffusion process approximately
isotropizes the electron distribution in the downstream fluid frame,
this is technically the reference frame for which the radiation modules
apply.  Therefore the applicable boost would be by a Lorentz factor
\teq{\Gamma \sim \Gamma_{\rm jet}} that is mildly-relativistic relative
to the mean jet motion represented by \teq{\Gamma_{\rm jet}}.  For
simplicity, here we ignore this subtle distinction to reduce the number
of model parameters, and set \teq{\Gamma\to \Gamma_{\rm jet}}.

To complete the radiation modeling, the gamma-ray portion of the
spectrum was then modified to treat line-of-sight attenuation of photons
in propagation from source to Earth.  This absorption derives from
\teq{\gamma\gamma\to e^+e^-} pair opacity due to the intervening
extragalactic background light (EBL) in the optical and infra-red; this
is of stellar and dust origin in galaxies. Such \teq{\gamma\gamma}
opacity due to the EBL has been extensively studied over the past two
decades, and many models for its space density and its impact on VHE
gamma-ray spectra of blazars and gamma-ray bursts exist. Here, we adopt
the attenuation correction model of \citet{FRD10}. Note that the
radiation models also treat \teq{\gamma\gamma\to e^+e^-} pair opacity
internal to the blazar jets, which can lead to attenuation of
$\gamma$-rays by IR and optical synchrotron fields.  In practice, the
jet Lorentz factors \teq{\Gamma_{\rm jet}} are chosen sufficiently high
for all three blazars studied that the jets are internally transparent
to this pair opacity.

Opting for a {\bf one-zone radiative model} here is motivated by
simplicity. In reality, one expects multi-zone structure to the emission
region, encompassing gradients in the field magnitude, electron density
and other system parameters.  Such complexity is the natural inference
from high-resolution imaging of jets in radio, optical and X-ray
wavebands.  Multi-zone constructions effectively introduce more modeling
variables, not necessarily precipitating a gain in insight.  The most
noticable change introduced by upgrading to a multi-zone scheme is the
broadening of the \teq{\nu F_{\nu}} peaks in both the synchrotron and
inverse Compton components: this muting of the spectral curvature of
these turnovers is a consequence of distributing the cooling rates for
each of the processes over a broader range of values.

The spatial structure conceived in a multi-zone model is a sequence of
segments subdividing the jet where acceleration and emission are
produced in each segment.  For our one-zone construction, we consider
just one such segment.  The acceleration zone is the most confined
portion of this region, being a planar slab of thickness \teq{2R_{\rm
acc}} with a planar shock dividing the zone roughly in halves.  In terms
of the acceleration, since the flow is relativistic, most of the
diffusive transport occurs in the downstream region, while the shock
drift contribution straddles the shock and encroaches primarily into the
upstream region.

\begin{figure}
\vspace*{-10pt}
\centerline{\includegraphics[width=9.0cm]{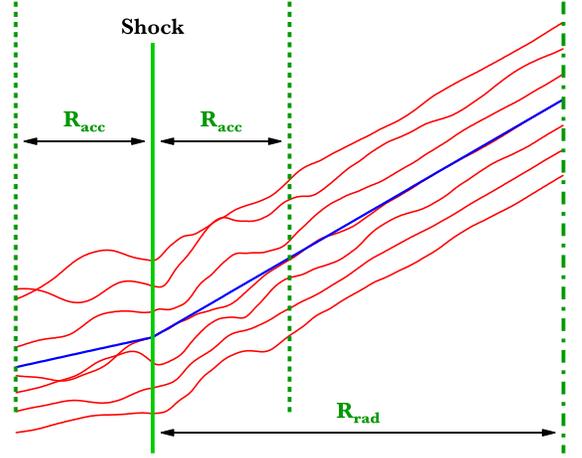}}
\vspace*{-20pt}
\caption{Schematic of the blazar model geometry under consideration, 
consisting of a region proximate to the shock that is the acceleration zone 
within a slab of approximate thickness \teq{2R_{\rm acc}}, which is 
embedded in a much larger radiation zone of size \teq{R_{\rm rad}} 
(not to scale).  The MHD background magnetic field structure of the shock 
is indicated by the blue straight line segments.  Superposed on this is 
a turbulent field, signified by the red field line projections that are 
computed from a Kolmogorov power spectrum of finite inertial range 
spanning around a decade in wavenumber \teq{k}.  This perturbation 
is exponentially damped on a scale of \teq{R_{\rm acc}} in this depiction 
so as to highlight confinement of turbulence to the shock layer. 
\label{fig:model_geom}}
\end{figure}

On larger scales, there is a radiation volume whose linear dimension
\teq{R_{\rm rad}} perpendicular to the shock layer is defined by the
approximate equality between the radiative cooling time (synchrotron
and/or inverse Compton, whichever is the shortest), and the jet fluid
convection time. Since the field is compressed by the shock, the
downstream region is where the synchrotron emission is most prevalent,
and if the gamma-rays are generated by the SSC mechanism, then here also
is where the inverse Compton signal is the strongest. MHD turbulence
will permeate the entire radiating volume, but is most likely more
concentrated near the shock, i.e. in the acceleration zone.  The reason
for this is that turbulence is naturally generated in the shock layer by
dissipation of the ballistic kinetic energy of the upstream flow as it
decelerates, and this turbulence should abate with distance from the
shock due to damping.  Such is the nature of interplanetary shocks and
their environs embedded in the solar wind.   A schematic of this model
geometry is given in Figure~\ref{fig:model_geom}, wherein the depicted
turbulent fields are actually a two-dimensional projection of
Alfv\'en-like fluctuations simulated using a Kolmogorov power spectrum
of modest inertial range.  An exponential damping of the fluctuations is
imposed on scales \teq{\vert x\vert \gtrsim 5} for the purposes of
illustration.

In the absence of information on the physical scale of the acceleration 
zone, the Monte Carlo simulations do not describe the high-energy 
cut-off of the electron distribution. We therefore extend the initial 
electron spectra like those in Fig.~\ref{fig:accel_dist} out to a maximum 
energy \teq{\gamma_{\rm max}mc^2 \approx \pmax c}.  Such an 
extrapolation is acceptable because the asymptotic power-law has 
been realized in the Monte Carlo simulation results with the displayed 
dynamic range in momenta.  This maximum energy is constrained in 
two ways: 
\vspace{-0pt}
\begin{enumerate}
\item particles will not be accelerated beyond an
energy for which the (synchrotron plus inverse Compton) radiative
cooling time scale, \teq{t_{\rm rad} = 3 m_e c^2 / (4\gamma\, c \, \sigt \, {\cal
U} )}, where \teq{{\cal U} = \UB + {\cal U}_{\rm rad}}
is the sum of the jet frame energy densities in the magnetic field and the photon
field, is shorter than the acceleration time scale, \teq{t_{\rm acc} =
\gamma m_e c \, \eta( p) / (e B)}, which will be discussed in greater
detail below;
\item particles will not continue to be accelerated once their energy 
establishes a diffusive mean free path
\teq{\lambda_{\parallel} = \eta(p) r_g(\gamma)\equiv \eta_1\,
(p/p_1)^{\alpha}} in Eq.~(\ref{eq:mfp_alpha}) that exceeds the size
\teq{R_{\rm acc}} of the acceleration zone.  Such energetic charges 
are assumed to escape upstream or downstream from the shock environs.
\end{enumerate}
Here \teq{\gamma = \{ 1+ (p/mc)^2 \}^{1/2} \approx p/(mc)} for energetic
leptons. With \teq{\gamma_{\rm max}} determined as the smaller of the
limiting values from the two constraints above, the high-energy portion
of the electron distribution {\it injected} at the acceleration site can
be written as \teq{n_{\rm acc} (\gamma) \propto \gamma^{-\sigma (p )} \,
\exp\left(-\gamma / \gamma_{\rm max} \right)} for \teq{\gamma\gg 1},
where \teq{\sigma} is the high-energy index of the {\it simulated} Monte
Carlo distribution \teq{dn/dp \propto p^{-\sigma}} of accelerated
particles.

The radiation module then uses as input an injected distribution, of
thermal particles plus charges accelerated up to \teq{\gamma_{\rm max}}
that are redistributed over the entire radiation zone, from which they
escape on a time scale \teq{t_{\rm esc} = \eta_{\rm esc}\, R_{\rm
rad}/c}. This escape parameter sets the normalization of the effective
equilibrium density of electrons in concert with the jet luminosity.  If
\teq{t_{\rm rad} <  t_{\rm esc}} when \teq{\gamma = \gammax}, a break in
the electron spectra is expected at a Lorentz factor 
\begin{equation}
   \gamma_{\rm br} \; \sim\; \dover{3 m_e c^2 }{4  \eta_{\rm esc}\, R_{\rm rad} \, \sigt \, {\cal U} } 
 \label{eq:gamma_br_cool}
\end{equation}
where the radiative cooling time equals the escape timescale. At this
\teq{\gamma}, the distribution steepens by an index increment
\teq{\Delta\sigma = 1}. As a reminder, \teq{{\cal U} = \UB + {\cal
U}_{\rm rad}} is the total field plus radiation energy density in the
co-moving frame of the jet, and this result applies for inverse Compton
scattering in the Thomson limit, a fairly representative situation. This
cooling break at \teq{\gamma_{\rm br} <\gamma_{\rm max}} is in addition
to acceleration flattening breaks illustrated in
Fig.~\ref{fig:accel_dist}, and generates a corresponding break in the
photon spectrum by an index of \teq{1/2}, the well-known signature of
the transition from slow to fast cooling at high Lorentz factors.

To couple the emission signal to the acceleration and cooling
information, the partially-cooled electron distributions are normalized
to a total kinetic luminosity of electrons in the blazar's two jets,
\begin{equation}
   L_{\rm kin} \; =\; 2\pi R_{\rm rad}^2 c \, \Gamma^2_{\rm jet} \, m_e c^2 
   \int_{1}^{\infty} n_e (\gamma) \, \gamma \, d\gamma
 \label{eq:Lkin}
\end{equation}
which is a free input parameter for fitting purposes.  The factor of 2
accounts for both the bright approaching jet that we observe, and the 
vastly fainter receding one.  As most of acceleration simulations illustrated 
in this paper produce distributions that are extremely flat, approaching 
\teq{n_e(\gamma )\propto \gamma^{-1}} (i.e., \teq{\sigma\sim 1}), 
a substantial and sometimes dominant contribution to \teq{L_{\rm kin}} 
comes from the non-thermal electrons up to the cooling break 
\teq{\gamma_{\rm br}}.  This is the case for the distributions exhibited 
in Fig.~\ref{fig:blazar_edist} below. Observe that since
changes in the electron spectrum and density will lead to changes in
the co-moving synchrotron photon field, our code determines the
equilibrium electron distribution through an iterative process.  It
starts by considering only the magnetic \teq{\UB} and external
radiation \teq{{\cal U}_{\rm rad}} energy densities to determine a
first-order equilibrium \teq{e^-} distribution. Then \teq{\UB} is
used to evaluate the co-moving synchrotron photon field, which is added
to the energy densities to re-determine the electron cooling rates and
to re-evaluate the break and maximum electron energies. The process is
repeated until convergence is achieved, in just a few iterations.

The magnetic field is specified by means of a magnetic partition
fraction \teq{\epsilonB \equiv \LB /L_{\rm kin}}, where 
\begin{equation}
   \LB\; =\; \pi R_{\rm rad}^2 \, c \, \Gamma_{\rm jet}^2 \, \UB
 \label{eq:mag_power_jet}
\end{equation}
is the power carried by the magnetic field along the jet, partly in the
form of Poynting Flux. Two powers of \teq{\Gamma_{\rm jet}} appear here,
defining the energy boost and the time dilation factors.  From the
perspective of the radiation modules, the field is effectively assumed
to be tangled in the co-moving jet frame so that isotropic emissivities
for both synchrotron and inverse Compton processes are employed.  In
reality, the acceleration considerations capture both turbulent and
quasi-laminar fields at different relative levels on different spatial
scales.  Furthermore, the shocks are moving at mildly-relativistic
speeds in the jet frame, thereby engendering ``beaming'' anisotropies in
the electron distribution; these will be neglected for the radiation
modeling since a number of the other parameters have a more important
impact on the results.  The radiative output from the equilibrium
electron distribution is evaluated using the radiation transfer modules
of \citet{Boettcher13}.

Two parameters derived from the spectroscopic models define the
characteristic rates for acceleration processes.  These are the
non-relativistic electron gyrofrequency, \teq{\omegaB}, and the electron
plasma frequency \teq{\omega_p}:
\begin{equation}
   \omegaB \; =\; \dover{eB}{m_ec}
   \quad ,\quad
   \omega_p \; =\; \sqrt{ \dover{4\pi n_e e^2}{m_e} }\quad .
 \label{eq:wB_wp}
\end{equation}
The gyrofrequency specifies the rate of acceleration in gyroresonant plasma 
mechanisms such as shock drift acceleration, while the plasma frequency 
governs the rate of energization in electrodynamic mechanisms such as 
magnetic reconnection.  The ratio of these two frequencies is captured 
in the non-relativistic plasma magnetization parameter
\begin{equation}
   \sigma \; =\; \dover{\omegaB^2}{\omega_p^2} 
   \; =\; \dover{B^2}{4\pi n_e m_ec^2} \quad ,
 \label{eq:sigma_param}
\end{equation}
familiar in kinetic plasma and MHD simulation studies.  The electron 
density \teq{n_e} is obtained from the kinetic luminosity \teq{L_{\rm kin}} 
in Eq.~(\ref{eq:Lkin}).  We highlight these parameters here as they 
will facilitate the interpretative elements later on.

\subsection{Generic Features of Synchrotron plus Inverse Compton Blazar One-Zone Models}

Before progressing to the blazar modeling, it is instructive to provide
an outline of the central elements of the multiwavelength fitting
diagnostics of the plasma environment in blazar jets.  The key
constraints derived are representative values of \teq{\eta (p) =
\lambda_{\parallel}/r_g} at the low injection momenta, \teq{p\sim
p_1\sim mu_1}, and at the maximum momentum \teq{\pmax} in the
acceleration zone.  These two mean free path parameters couple via the
index \teq{\alpha} of the diffusion law in Eq.~(\ref{eq:mfp_alpha}).

Foremost in our considerations is the value of \teq{\eta (\pmax )},
which far exceeds unity for our chosen blazars.  The origin of this
property is that large \teq{\eta} are required to fit the frequency of
the synchrotron \teq{\nu F_{\nu}} peak just below the spectral turnover
of this emission component. For strong cooling in the acceleration zone,
the turnover is created when the cooling rate
\begin{equation}
   -\, \dover{d\gamma}{dt}\biggl\vert_{\rm cool} \; =\; \dover{4 \sigt}{3m_ec} \, {\cal U} \, \gamma^2
 \label{eq:cool_rate}
\end{equation}
in the jet frame is approximately equal to the electron acceleration 
rate for the diffusive Fermi process
\begin{equation}
   \dover{d\gamma}{dt}\biggl\vert_{\rm acc} \;\sim\; \left( \dover{u_{1x}}{c} \right)^2
   \, \dover{eB}{\eta\, m_ec}
   \quad\hbox{for}\quad
   \eta (p) \; =\; \eta_1 \left( \dover{p}{p_1} \right)^{\alpha-1}\; .
 \label{eq:acc_rate}
\end{equation}
Note that this acceleration rate is an approximate scale, and more
precise forms that isolate upstream and downstream contributions, and
field obliquity influences can be found in reviews like \citet{Drury83}
and other papers such as \citet{Jokipii87}. Now introduce a cooling
parameter \teq{\epsilon_{\rm syn} = \UB/{\cal U}} that represents the
fractional contribution of the synchrotron process to the electron
cooling in the jet frame.  The contribution \teq{{\cal U}_{\rm rad}} to
\teq{{\cal U}} is determined from the comoving radiation field
(radio-to-gamma-ray) from the radiation codes as outlined above. This
synchrotron parameter can be approximately expressed in terms the ratio
of optical/X-ray ``peak luminosities'' to the pseudo-bolometric
luminosity: \teq{\epsilon_{\rm syn}\approx L_{O-X}/(L_{O-X} +
L_{\gamma})}. For Mrk 501, this has a value of about \teq{\epsilon_{\rm
syn} \sim 0.5-0.6}, as can be inferred from the broadband spectrum in
Fig.~\ref{fig:MW_spec_Mrk501}. In contrast, for the
synchrotron-dominated blazar BL Lacertae, \teq{\epsilon_{\rm syn} \sim
0.8-0.9}, which is somewhat higher than the values inferred for Mrk 421
and Mrk 501. For the flare state of AO 0235+164, the strong gamma-ray
signal sets \teq{\epsilon_{\rm syn} \sim 0.25}.

The acceleration/cooling equilibrium then establishes a maximum electron
Lorentz factor at
\begin{equation}
   \gammax \; =\; \left( \dover{9\epsilon_{\rm syn}}{4\eta_1} 
   \, \Bigl( \dover{u_{1x}}{c} \Bigr)^2 \, \dover{e}{B\, r_0^2} \right)^{1/(1 + \alpha )} \quad ,
 \label{eq:gamma_max}
\end{equation}
assuming that \teq{p_1\sim m_ec} in a mildly relativistic shock with 
\teq{u_{1x}\sim c}.  Here \teq{r_0 = e^2/m_ec^2} is the classical 
electron radius.  If \teq{\epsilon_{\rm syn}\sim 1}, then synchrotron 
emission dominates the cooling of particles, and when \teq{\alpha =1},
corresponding to \teq{\eta} being independent of electron momentum, 
the well-known result that \teq{\gammax \propto B^{-1/2}} follows.  
Since \teq{e/Br_0^2 = B_{\rm cr}/(\fsc B)} for a fine structure constant 
\teq{\fsc =e^2/(\hbar c)} and \teq{B_{\rm cr} = m_e^2 c^3/e\hbar \approx 
4.41\times 10^{13}}Gauss as the Schwinger field, the scale of \teq{\gammax} 
in the range \teq{\sim 10^3-10^6} is established for blazars with fields 
\teq{B\sim 1-10}Gauss in the jet frame, depending on the value of \teq{\alpha}.  
This is readily seen by recasting Eq.~(\ref{eq:gamma_max}) in the form
\begin{equation}
   \gammax \; \sim\; \sqrt{ \dover{2 {\cal E}_s(\alpha )}{3} } 
   \left(  \dover{6\times 10^{15}}{\eta_1\, B} \right)^{1/(1 + \alpha )} \quad ,
 \label{eq:gammax_syn}
\end{equation}
for \teq{B} in units of Gauss, and
\begin{equation}
   {\cal E}_s(\alpha ) \; =\; \dover{3}{2} \left( \dover{9 \epsilon_{\rm syn}}{4} 
   \, \Bigl[ \dover{u_{1x}}{c} \Bigr]^2 \right)^{2/(1 + \alpha )} 
 \label{eq:calE_s_def}
\end{equation}
being of the order of unity for blazar shocks.  Increasing \teq{\alpha} 
above unity lowers \teq{\gammax} because the acceleration 
becomes less efficient (slower) in competing with cooling.

The synchrotron turnover/cutoff energy \teq{E_{\rm syn}\propto \gammax^2 B}
corresponding to \teq{\gammax} is 
\begin{equation}
  E_{\rm syn} \; \sim\; \dover{\delta_{\rm jet}\,  {\cal E}_s(\alpha )}{\eta_1\, (1+z)}  \,
   \left( \dover{\fsc B}{B_{\rm cr}} \eta_1 \right)^{(\alpha -1)/(\alpha + 1)} \dover{m_ec^2}{\fsc} \quad ,
 \label{eq:Esyn_max}
\end{equation}
The blueshift factor \teq{\delta_{\rm jet} = 1/[\Gamma_{\rm jet}(
1-\beta_{\rm jet}\cos\theta_{\rm obs} )]} due to Doppler beaming has
been included, where \teq{\theta_{\rm obs}\ll 1} is the observer's
viewing angle with respect to the bulk velocity of the jet in the
emission region.  Also, \teq{z} is the redshift of the distant blazar,
so that \teq{E_{\rm syn}} represents an observer's measurement of the
energy.  For the Bohm limiting case of \teq{\eta (\gammax )=1}, imposing
both \teq{\alpha=1} and \teq{\eta_1=1}, and setting \teq{\delta_{\rm
jet} = 1}, the \teq{\Gamma_1 =1} and \teq{u_{1x}=c} evaluation yields
\teq{E_{\rm syn}\sim 100} MeV, independent of the magnetic field
strength.  This widely-known result was highlighted in \citet{DeJager96}
for considerations of $\gamma$-ray emission at relativistic pulsar wind
nebular shocks, but was derived much earlier by \citet{GFR83} in the
context of AGN. This special result holds provided that the acceleration
process is gyroresonant, which is the prevailing paradigm for both
non-relativistic and relativistic shocks.

For Mrk 501 and our other case study blazars, \teq{\gammax} becomes
weakly dependent on \teq{B}, and the turnover energy \teq{E_{\rm syn}}
quickly drops below the \teq{m_ec^2/\fsc} scale. For representative jet
fields \teq{B\sim 1}Gauss, one quickly estimates using
Eq.~(\ref{eq:Esyn_max}) that for the synchrotron \teq{\nu F_{\nu}} peak
energy \teq{E_{\rm syn}} to appear in the optical, one requires values
\teq{\alpha \sim 2.5-3}, and often also values \teq{\eta_1\gtrsim 100}.
This is a central feature of our case studies below: in order to move
the synchrotron peak into either the X-ray or optical band, large values
of \teq{\eta (\gammax )} are required. In such circumstances, the
acceleration process may not be gyroresonant, a realization that we will
discuss in Sec.~\ref{sec:discussion}. The compelling need for \teq{\eta
(\gammax )\gg 1} in the shock acceleration interpretation of blazar
activation was first emphasized by \citet{IT96}, who chose a {\it
momentum-independent} \teq{\eta \sim 10^5} when exploring
multi-wavelength modeling of Mrk 421 spectra. In our more versatile
construct here, \teq{\alpha} is not fixed to unity. To realize such
large \teq{\eta} at \teq{\gammax}, without forcing \teq{\eta_1} to
similar large values that would suppress efficient injection of charges
from the thermal population into the Fermi mechanism \citep{SB12},
values \teq{\alpha > 1} are necessitated.

Synchrotron self-Compton models possess an additional constraint on
system parameters via the observer's frame energy of the SSC peak
\teq{\ESSC} in the hard gamma-rays.  This is just positioned according
to the energy gained by synchrotron photons of around the turnover
energy \teq{E_{\rm syn}} subjected to inverse Compton scattering.  For
the Thomson limit, this energy enhancement ratio is \teq{4\gammax^2/3},
so that
\begin{equation}
   \gammax \;\sim\; \sqrt{\dover{3 \ESSC}{4E_{\rm syn}} } \quad .
 \label{eq:gammax_SSC}
\end{equation}
In conjunction with Eq.~(\ref{eq:Esyn_max}), this restricts the values
of \teq{B} and \teq{\alpha}.  This constraint on \teq{\gammax} is not
applicable for external inverse Compton models where the target photon
field is not the synchrotron population, but perhaps originating from a
proximate disk; this will actually be the case for our blazars BL Lac
and AO 0235+164. It is also modified somewhat when the upscattering
samples the Klein-Nishina regime, a domain that is barely encroached
upon in most of our models, since they satisfy \teq{4\gammax E_{\rm
syn}/(\delta_{\rm jet}\, m_ec^2) \lesssim 1}.

\begin{figure}
\vspace*{-10pt}
\centerline{\hskip 10pt\includegraphics[width=9.5cm]{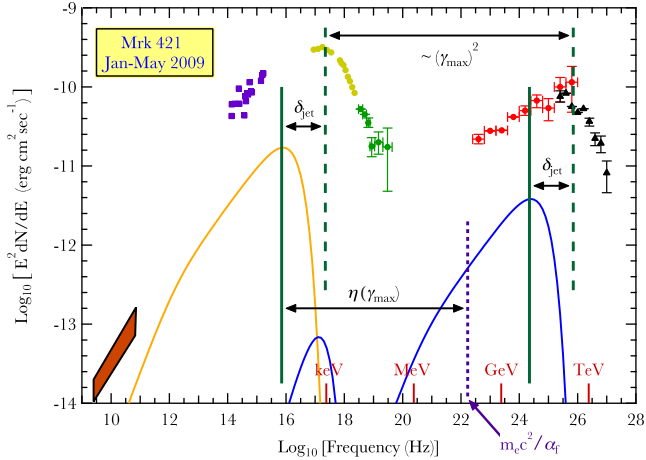}}
\vspace*{-5pt}
\caption{
Generic construction of multiwavelength synchrotron self-Compton (SSC)
spectral modeling for blazars in the \teq{\nu F_{\nu}} representation.
The illustration includes a selection of data for Mrk 421 published in
Fig.~8 of \citet{Abdo11b} for a campaign during the first half of 2009. 
This includes the ochre band signifying various radio detections, purple
squares for optical data from different facilities, yellow-green
(Swift-XRT) and green circles (Swift-BAT) for X-rays, red dots for {\it
Fermi}-LAT measurements, and black triangles for MAGIC TeV-band data.
The solid curves are the schematic model spectra applicable to the
comoving jet frame, with radio-to-optical/UV (orange) curve constituting
the synchrotron component, and the X-ray-to-\teq{\gamma}-ray (blue)
curves denoting the SSC spectrum. The frequency offsets between the
model peak and the data peak for these two components (marked by solid
and dashed green vertical lines, respectively) approximately define the
value of the jet Doppler factor \teq{\delta_{\rm jet}}. The separation
of the peaks of the synchrotron and SSC components is by a factor of the
order of \teq{\gammax^2}.  In the jet frame, the synchrotron model peak
lies at an energy that is a factor \teq{\eta (\gammax )} below the
fundamental bound of \teq{m_ec^2/\fsc} (see text).
\label{fig:spec_schematic}}
\end{figure}

This ensemble of diagnostics for our model parameters is summarized in
Fig.~\ref{fig:spec_schematic}.  In it, an array of observational data
from the multiwavelength campaign on Mrk 421 of January -- May 2009 is
depicted: this serves as a selection from Fig.~8 in \citet{Abdo11b}. Mrk
421 was chosen for the Figure as it will not be studied here, though we
note that its broadband spectrum is qualitatively similar to that of our
case study blazar Mrk 501.  Schematic model synchrotron and SSC spectra
are exhibited in Fig.~\ref{fig:spec_schematic}, though specifically as
would be computed in the jet rest frame.  This then permits
identification of the Doppler factor \teq{\delta_{\rm jet}} via
measurement of the frequency ratio between peaks in the observed and
modeled spectra for each of the emission components.  Additionally, the
ratio of observed to theoretical peak heights measures the flux
enhancement \teq{\delta_{\rm jet}^4} between model and data, noting that
this element of the schematic is not to scale. The separation of the
component peaks (observed or modeled) defines the inverse Compton
scattering enhancement factor \teq{4\gammax^2/3} (Thomson limit), and so
constrains the value of \teq{\gammax} approximately according to
Eq.~(\ref{eq:gammax_SSC}).  For the illustrated case, \teq{\gammax\sim
10^4} and the synchrotron peak energy is around \teq{10^{-4}m_ec^2}, so
that the Compton scattering near the \teq{\sim}TeV peak is marginally in
the Klein-Nishina regime. These are standard paths to computing spectral
fitting parameters for SSC models of blazars. What is more unique to the
present study is the determination of the diffusion parameter \teq{\eta
(\gammax )} by comparing the synchrotron model peak energy with the
fundamental bound \teq{m_ec^2/\fsc}, as depicted in
Fig.~\ref{fig:spec_schematic}.  This is essentially inverting
Eq.~(\ref{eq:Esyn_max}) to yield a combination of \teq{\eta_1} and
\teq{\alpha} that is captured in Eq.~(\ref{eq:eta1_alpha_reln}) below. 
These two diffusion parameters will form a central focus for our case
studies below.

Observe the appearance of a thermal SSC component in the X-rays due to
bulk Comptonization of shock-heated thermal electrons; this actually
emerges smoothly above the non-thermal inverse Compton contribution at
flux levels below those of the plot scale.  A similar thermal
synchrotron component is not exhibited (below 1 GHz) in the Figure,
since it would be suppressed by synchrotron self-absorption in the
radio. Each component also exhibits a cooling break, at \teq{E_{\rm
br,syn} \sim 3\times 10^{12}}Hz and \teq{E_{\rm br,SSC} \sim 3\times
10^{21}}Hz, respectively, where the photon spectrum steepens by an index
of \teq{1/2} at higher frequencies.  The approximate value of
\teq{\gamma_{\rm br}} can be deduced since \teq{\gamma_{\rm br}^2}
roughly represents the ratio of the energy of the break to the thermal
peak energy in the SSC component.  This follows from the presumption
that the shock heated electrons are at most mildly-relativistic (true
for mildly-relativistic blazar shocks), and one would conclude that
\teq{\gamma_{\rm br}\sim 10^2} from the depiction in
Fig.~\ref{fig:spec_schematic}. The precise value of \teq{\gamma_{\rm
br}} in Eq.~(\ref{eq:gamma_br_cool}) yields a measure of the relative
sizes of the acceleration and cooling zones.

It is evident that constraints imposed by the energies of the
synchrotron and SSC \teq{\nu F_{\nu}} peaks on jet environmental
parameters lead to strong couplings between them. The most obvious of
these is the calibration of the magnetic field strength using the ratio
of the SSC and synchrotron peak frequencies.  Combining
Eqs.~(\ref{eq:gammax_syn}) and~(\ref{eq:gammax_SSC}), one arrives at the
coupling between \teq{B} and \teq{\alpha}:
\begin{equation}
   B \;\sim\; \dover{6\times 10^{15}}{\eta_1} 
   \left( \dover{8E_{\rm syn}}{9 \ESSC} \right)^{(1 + \alpha )/2}
 \label{eq:B3_SSC_syn_ratio}
\end{equation}
for \teq{{\cal E}_s(\alpha )\sim 1}.  Evidently, for synchrotron peaks
in the X-rays (HBLs), fields \teq{B\sim 1}Gauss require \teq{\alpha \sim
1-1.5}, whereas, for blazars with synchrotron peaking in the optical,
higher values of \teq{\alpha} are needed to establish similar jet
fields.  The ratio of the SSC to synchrotron peak luminosities also
provides a modest constraint on \teq{B} via the \teq{\epsilon_{\rm syn}}
parameter.

The second diffusion/acceleration parameter, \teq{\eta_1}, can be
introduced by inverting Eq.~(\ref{eq:Esyn_max}), again for the case
\teq{{\cal E}_s(\alpha )\sim 1}.  Set \teq{\nu_{\hbox{\fiverm S,14}}} to
be the synchrotron peak frequency in the comoving frame in units of
\teq{10^{14}}Hz, and \teq{\nu_{\hbox{\fiverm SSC,24}}} as the SSC peak
frequency in the jet frame in units of \teq{10^{24}}Hz.  Eliminating
\teq{B} using Eq.~(\ref{eq:B3_SSC_syn_ratio}), one can then ascertain
the fiducial relationship between \teq{\eta_1} and \teq{\alpha} for
representative observed blazar synchrotron and SSC peak energies:
\begin{equation}
   \eta_1^{2/(\alpha +1)} \;\sim\; \dover{1.5 \times 10^{-2} 
   \delta_{\rm jet}}{(1+z)\, \nu_{\hbox{\fiverm SSC,24}}} 
   \, \left( 10^{-10}\, \dover{\nu_{\hbox{\fiverm S,14}} }{\nu_{\hbox{\fiverm SSC,24}} } \right)^{(\alpha -3 )/2}\quad .
 \label{eq:eta1_alpha_reln}
\end{equation}
With \teq{\delta_{\rm jet}\sim 10-30}, this clearly suggests that 
\teq{\alpha >2} cases are needed for synchrotron peaks to appear 
in the optical, and concomitantly \teq{\eta_1 \gtrsim 30-100}.  The 
physical implications of such diffusion parameters will be addressed 
in Section~\ref{sec:discussion}. Lower values of \teq{\alpha} and 
\teq{\eta_1} can be entertained if the synchrotron peak appears 
in the X-rays, as it does for Mrk 501.

\subsection{Case Study: Mrk 501}

An obvious first choice for study is the well-monitored BL Lac blazar
Markarian 501, located at a redshift of \teq{z=0.034}.  Multiwavelength
campaigns are readily available for this source
\citep[e.g.,][]{Catanese97,Albert07}, thereby eliminating problems
associated with non-contemporaneous data in different wavebands.  The
more recent campaign conducted during the March--July 2009 epoch serves
amply to illustrate the advances in understanding offered by combining
shock acceleration simulation results with radiation emission codes. The
broadband spectral character of Mrk 501 differs only modestly from that
depicted in Fig~\ref{fig:spec_schematic} for its sister blazar Mrk 421,
and is also fairly similar to the SED of more distant blazars such as
PKS 2155-304 \citep[at \teq{z=0.116}; see][]{Aharonian09}, yet it is
still suitable to an SSC model construction with an added optical
component.

\begin{figure}
\vspace*{-10pt}
\centerline{\hskip 10pt\includegraphics[width=9.5cm]{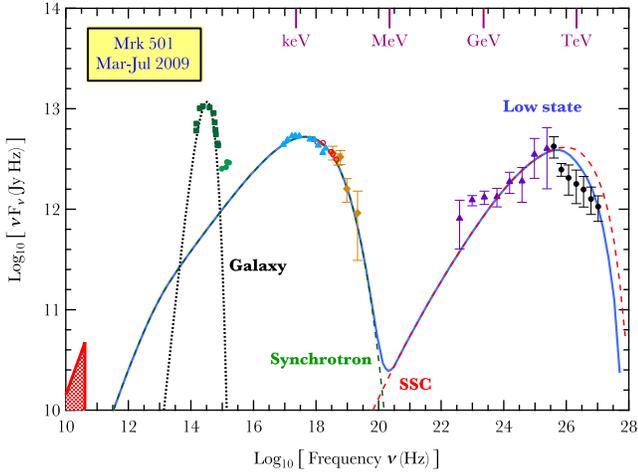}}
\vspace*{-5pt}
\caption{Multiwavelength \teq{\nu F_{\nu}} spectra (points), together with model
fits as described in the text, for the extended March-July 2009
monitoring of the blazar Markarian 501. The campaign data are taken from
Figures 8 and 9 of Abdo et al. (2011), and constitute a ``low'' state
with gamma-ray data bracketing the flare of early May, 2009 being
omitted (see text). The gamma-ray detections and upper limits are from
{\it Fermi}-LAT (purple triangles) and MAGIC (black circles), and the
X-ray points are {\it Swift}-BAT (muddy orange squares), {\it Swift}-XRT
(light blue triangles) and RXTE-PCA data (open red circles).  Optical,
UV and radio measurements are detailed in Abdo et al. (2011); since a
variety of radio fluxes were recorded for various regions much larger
than the compact X-ray/$\gamma$-ray zones, they are marked by a
representative band, and were not fit. The broadband models consist of a
synchrotron component (dashed green curve) up to the X-ray band, an SSC
component in the X-rays and gamma-rays (dashed red curve), and a
separate thermal host galaxy emission component (dotted black). The full
jet model spectra for the extended low state includes a correction \citep{FRD10} 
for \teq{\gamma\gamma} absorption by the extragalactic background light,
and combined with the data, constrain the diffusion parameters to
\teq{\eta_1=100} and \teq{\alpha =1.5} (see text).
\label{fig:MW_spec_Mrk501}}
\end{figure}

An extensive summary of the mid-March--early August, 2009 campaign
involving the VERITAS and MAGIC telescopes above 100 GeV, {\it
Fermi}-LAT in the 20MeV to 300 GeV energy window, RXTE-PCA (2--60 keV)
and Swift XRT+BAT (0.3-150 keV) in the X-rays, combined with UV
(Swift-UVOT), optical and radio measurements, is provided in
\citet{Abdo11a}. For most of this epoch, the source was in a
low/moderate state, for which broadband spectral data are depicted in
Fig.~\ref{fig:MW_spec_Mrk501}.  During this campaign, in the short
window May 1--5, 2009, Mrk 501 underwent a sizable flare in gamma-rays,
evincing a high state at around 5 times the Crab nebula flux  in the
VERITAS/MAGIC/{\it Fermi}-LAT bands.  This was an enhancement of a
factor of 3-5 over the quieter portions of the campaign, and received
considerable attention in \citet{Abdo11a} and also \citet{Aliuetal16}. 
The flare exhibited a spectrum that was much harder above
\teq{10^{23}}Hz than for the extended state. To aid clarity, the VERITAS
data were not depicted in Fig.~\ref{fig:MW_spec_Mrk501}, being similar
to the MAGIC points above \teq{10^{25}}Hz.  Moreover, the
multi-wavelength data represent the long-term average over the 4.5 month
interval, and for radio, optical, UV and X-rays, were taken from the
spectrum in Fig.~8 of \citet{Abdo11a}.  The gamma-ray points, for {\it
Fermi}-LAT and MAGIC, omit an interval of 30 days that brackets the
strong flaring episode, and are taken from Fig.~9 of \citet{Abdo11a}. 
Note that the units for the $y$-axis in this and subsequent
multi-wavelength spectral figures employ the Jansky-Hertz choice
familiar to radio astronomers, and can be converted to the form in
Fig.~\ref{fig:spec_schematic} that is often used by high energy
astrophysicists via 1 Jy Hz \teq{=10^{-23}} erg cm$^{-2}$ sec$^{-1}$.

In the SSC interpretation, the extended state is very slightly
synchrotron-dominated, contrasting the May 1--5, 2009 high state, which
is marginally inverse Compton-dominated; throughout, the synchrotron
peak is positioned in the X-ray band.  The low-variability UV/optical
data is not attributed exclusively to the jet, but mostly to the host
galaxy: at least 2/3 of the flux is believed
\citep[e.g.,][]{Nilssonetal07} to originate in the host, and is herein
modeled via a separate thermal component.  The radio data were from
various (mostly single-dish) facilities \citep[see the list in Table 1
of][]{Abdo11a} that generally do not resolve the jet, and accordingly
provide upper bounds to the signal expected from the highly-variable
$\gamma$-ray emission zone.  For this reason, they are depicted in
Fig.~\ref{fig:MW_spec_Mrk501} as a swath, positioned at the fluxes
measured at the various radio frequencies.

Using the radiation modeling protocols and coding described in
Section~\ref{sec:radmodels}, multiwavelength spectra from radio to VHE
gamma-rays were generated for a suite of particle distributions
generated by the Monte Carlo shock acceleration simulation.  The
turbulence parameters \teq{\alpha} and \teq{\eta_1} were adjusted for
the various simulation runs to hone in on a candidate ``best case,'' the
spectral results for which are exhibited in
Fig.~\ref{fig:MW_spec_Mrk501}. The spectra so generated are dominated by
two components, synchrotron emission in the radio-to-X-ray band, and a
synchrotron self-Compton signal in the gamma-ray range; the observer
frame computations of these components at the source are isolated in
this Figure as dashed curves.  The overall spectrum at Earth is
subjected to \teq{\gamma\gamma} attenuation from the line-of-sight EBL,
as described above, and is depicted as the solid M/W curve in
Fig.~\ref{fig:MW_spec_Mrk501} that is the nominal fit to the data; this
attenuation is omitted in the SSC component depiction so as to
illustrate its magnitude.

Simulation and radiation model input parameters, and also those derived
from the input ones, are listed in Table~1; some of them can be compared
with model fitting parameters listed in Table~2 of \citet{Abdo11a}.  
Since in this broadband spectrum the thermal and non-thermal components
are obviously distinct, the detailed shape of the host galaxy optical
spectrum is irrelevant for our modeling goals. Therefore, no attempt was
made to perform a detailed fit to the optical (e.g., with a spectral
template for an elliptical galaxy); this component was simply
represented with a separate blackbody spectrum for illustrative purposes.

\begin{table*}
 \centering
 \begin{minipage}{140mm}
  \caption{Input and Derived Parameters for Blazar Models}
  \begin{tabular}{@{}lrcccc@{}}
  \hline
   \multicolumn{2}{c}{Parameter}  & Mrk 501
        & BL Lacertae & AO 0235+164 \\
   Name & Symbol/units & Extended State & Extended State & High State \\
    \hline
  \multicolumn{5}{c}{Jet+Source Parameters} \\
  \hline
    Redshift & $z$ & 0.034 & 0.069 & 0.94 \\
    Jet Lorentz factor & $\Gamma_{\rm jet}$ & 25  & 15 & 35  \\
    Observing angle & $\theta_{\rm obs}$ & $2.29^{\circ}$  & $3.82^{\circ}$ & $1.7^{\circ}$  \\
    Emission region size & $R_{\rm rad}$ [cm] & $1.2\times 10^{16}$  & $2.5\times 10^{15}$ & $1.0\times 10^{16}$  \\
    $e^-$ injection luminosity\footnote{Expressed in Eq.~(\ref{eq:Lkin}), 
          wherein the electron number density $n_e$ is normalized.} & $L_{\rm kin}$ [erg/sec] 
          & $1.5\times 10^{44}$ & $6.7\times 10^{43}$ & $4.8\times 10^{46}$  \\
    Escape time scale\footnote{The escape time is given by \teq{t_{\rm esc} = \eta_{\rm esc}\, R_{\rm rad}/c}.} 
         & $\eta_{\rm esc}$ & 100 & 5 & 300  \\
    Magnetic partition & \teq{\epsilonB \equiv \LB /L_{\rm kin}} & $3\times 10^{-4}$ & $0.5$ & 0.06  \\
    Dusty torus luminosity & [erg/sec] & --- & $6\times 10^{44}$ & $3.4\times 10^{44}$ \\
    Dusty torus temperature & [K] & --- & $2\times 10^3$ & $10^3$ \\
    \hline
    \multicolumn{5}{c}{Shock Parameters} \\
    \hline 
    Upstream speed & $u_{1x}/c$ & 0.71 & 0.71 & 0.71  \\
    Field obliquity & $\ThetaBfone $ & $32.3^{\circ}$ & $32.3^{\circ}$ & $52.4^{\circ}$  \\
    Upstream temperature & $T_1$ (K) & $5.45\times 10^7$ & $5.45\times 10^7$ & $5.45\times 10^7$  \\
    Compression ratio & $r$ & 3.71 & 3.71 & 3.71  \\
    Injection mean free path & $\eta_1=\eta(p_1)$ & 100 & 20 & 225  \\
    Diffusion index & $\alpha$ & 1.5 & 3 & 3  \\
    \hline
    \multicolumn{5}{c}{Derived Parameters} \\
    \hline 
    Magnetic field & $B$ [Gauss] & $1.15\times 10^{-2}$ & $2.52$ & $2.50$  \\
    Field luminosity & $\LB$ [erg/sec] & $4.5\times 10^{40}$ & $3.35\times 10^{43}$ & $2.9\times 10^{45}$   \\
    Synchrotron partition\footnote{This is given by \teq{\epsilon_{\rm syn}\approx L_{O-X}/(L_{O-X} + L_{\gamma})}.} 
         & $\epsilon_{\rm syn}$ & 0.6 & 0.83 & 0.25  \\
    Cyclotron frequency & $\omegaB$ & $2.0 \times 10^5$ & $4.44\times 10^7$ & $4.4\times 10^7$  \\
    $e^-$ plasma frequency & $\omega_p$ & $1.95\times 10^5$ & $1.4\times 10^7$ & $4.0\times 10^7$  \\
    Shock magnetization\footnote{Eq.~(\ref{eq:sigma_param}) applies to non-relativistic cases; it must be
             divided by the Lorentz factor $\Gamma_1$ for relativistic shocks.} & $\sigma$ & $1.08$ & $10.0$ & $1.2$  \\
    Alfv\'en speed & $ B/\sqrt{4\pi \rho} $ & $2.34\times10^{-3}\,$c & $0.074\,$c & $0.026\,$c \\ 
    Cooling break & $\gamma_{\rm br}$ & $3.2\times 10^3$ & $1.75\times 10^2$ & 2.0  \\
    Maximum $e^-$ Lorentz factor & $\gamma_{\rm max}$ & $8.5\times 10^5$ & $3.64\times 10^3$ & $1.61\times 10^3$  \\
    Maximum $\lambda_{\parallel}/r_g$ & $\eta (\gammax )$ & $9.2 \times 10^4$ & $2.6 \times 10^8$ & $5.8 \times 10^8$  \\
    Maximum mean free path\footnote{This is $\lambda_{\parallel}(\gammax) = \gammax\eta (\gammax )\, c/\omegaB$, and is realized
         in the acceleration region and beyond.} & $\lambda_{\rm max}$ [cm] & $1.17 \times 10^{16}$ & $6.5 \times 10^{14}$ & $6.4 \times 10^{14}$  \\
    Synchrotron cutoff & $\nu_{\rm syn}$ [Hz] & $4.2 \times 10^{17}$ & $3.1 \times 10^{14}$ & $1.7 \times 10^{14}$  \\
    SSC cutoff frequency & $\nu_{\hbox{\fiverm SSC}}$ [Hz] & $8.8 \times 10^{25}$  & $7.2 \times 10^{20}$ & $5.9 \times 10^{20}$  \\
    \hline
\end{tabular}
\end{minipage}
\end{table*}

% ne = Lkin/( Pi Rrad^2 c Gamjet^2 me c^2 10.0) /. {Lkin -> 1.5 10^44,  Gamjet -> 25, Rrad -> 1.2 10^16}
% ne = Lkin/( Pi Rrad^2 c Gamjet^2 me c^2 10.0) /. {Lkin -> 6.7 10^43,  Gamjet -> 15, Rrad -> 2.5 10^15}
% ne = Lkin/( Pi Rrad^2 c Gamjet^2 me c^2 10.0) /. {Lkin -> 4.8 10^46,  Gamjet -> 35, Rrad -> 1.0 10^16}

The principal constraints derived from the Mrk 501 multiwavelength
modeling are that the mean free path parameter \teq{\eta\sim \eta_1} is
modest for mildly-relativistic electrons injected into the shock
acceleration process, and that at the highest energies, \teq{\eta
(\gammax)\gtrsim 10^5} in order to place the synchrotron turnover (i.e.,
\teq{\nu F_{\nu}} peak) in the hard X-ray band.  The values we obtained
for the extended state were \teq{\eta_1=100} and \teq{\alpha =3/2} (i.e.
\teq{\eta (p) \propto p^{1/2}}). These parameters generate moderately
efficient injection from thermal energies into the Fermi mechanism (see
Fig.~\ref{fig:accel_dist}), and mean free paths that possess a stronger
momentum dependence than Bohm diffusion (\teq{\alpha =1}), suggesting
interactions with weaker turbulence for more energetic particles
diffusing on larger spatial scales.  The interpretive elements of this
parametric fit will be embellished upon in Section~\ref{sec:discussion}.

It should be borne in mind that this fitting ``solution'' is
representative of the appropriate parameter space, but is by no means a
unique choice.  It can be generally stated that there is a tolerance of
about \teq{\pm 0.2} in \teq{\alpha}, and a tolerance of a factor of
\teq{\sim 1-5} in \teq{\eta_1} permitting M/W spectral fits of similar
quality.  The uncertainties in the gamma-ray fluxes, particularly in the
{\it Fermi}-LAT band, preclude greater precision for modeling. 
Moreover, in this analysis, the shock MHD parameters were kept fixed at
the values chosen for Figure~10 of \citet{SB12}, namely a shock speed of
\teq{u_{1x}=0.71c}, a velocity compression ratio of \teq{r=2.71}, a
field obliquity of \teq{\ThetaBfone = 32.3^{\circ}}, and a low sonic
Mach number of around 4. Adjusting the MHD structure of the shock would
introduce changes to the optimal choices for the shock layer turbulence
parameters \teq{\alpha} and \teq{\eta_1} in a data fitting protocol, so
the ones for spectra exhibited in the Figure serve as general guides,
and should not be presumed sacrosanct.

Likewise, the choice of the observational epoch should always be borne
in mind: Mrk 501 is highly variable above the optical band, and so sets
of fitting parameters will change when different epochs are modeled. 
This is evident for strong flaring episodes such as the May 1--5, 2009
event not treated here.  Yet it also applies for other observational
campaigns, such as the recent NuSTAR-led multi-wavelength monitoring
spanning the period April 1 -- August 10, 2013 \citep{Furniss15}.  The
X-ray spectrum therein is often harder and more luminous than that
depicted here, though the gamma-ray spectrum (MAGIC+VERITAS+{\it
Fermi}-LAT) is more or less commensurate with that in
Fig~\ref{fig:MW_spec_Mrk501} --- see Fig.~9 of \citet{Furniss15} for
details.  To appreciate how extreme Mrk 501 can become during flares,
one need look no further than the depiction of the April 1997 flare in
Fig.~6 of \citet{Acciari11}, wherein the synchrotron peak moves to above
\teq{3\times 10^{19}}Hz (i.e. nearly two decades higher than the listing
in Table~1), and its flux level is slightly over a decade higher than
that illustrated in Fig.~\ref{fig:MW_spec_Mrk501}. Modeling this extreme
HBL behavior would lower the inferred value of \teq{\eta (\gammax )} by
a factor of 30--50.  Yet for this flare, and also for the NuSTAR M/W
campaign, the principal conclusion of a strong momentum dependence for
the diffusion coefficient and large departures from the Bohm limit at
\teq{\gamma\sim\gammax} would be upheld, and is anticipated to apply to
a broad selection of observing campaigns.

\begin{figure}
\vspace*{-10pt}
\centerline{\hskip 10pt\includegraphics[width=9.3cm]{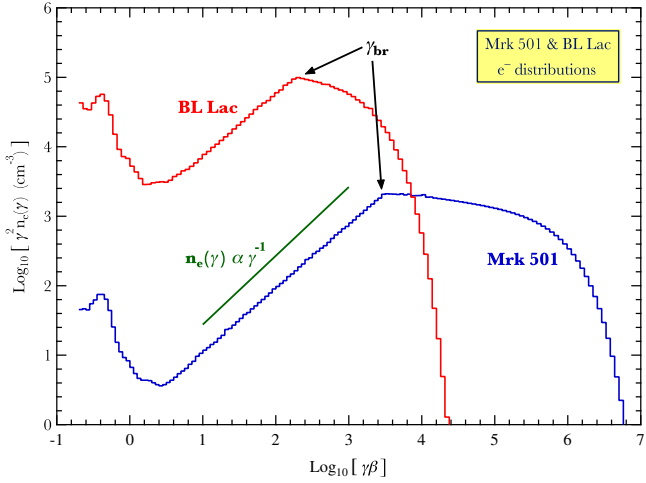}}
\vspace*{-5pt}
\caption{Complete thermal plus non-thermal electron distributions employed 
in the multiwavelength modeling results for Mrk 501 depicted in 
Fig.~\ref{fig:MW_spec_Mrk501}, and BL Lacertae as shown in 
Fig.~\ref{fig:MW_spec_BLLac}, as functions of the four-velocity or 
dimensionless electron momentum \teq{\gamma\beta}.   These distributions, 
applicable to the jet frame, are derived from the Monte Carlo simulation 
results that are injected into the radiating volume of the blazar jet, and are 
modified by cooling losses above the cooling break energy 
\teq{\gamma_{\rm br}\, m_ec^2}, and eventually by the suppression of 
acceleration by rapid radiative cooling at \teq{\gamma\sim \gamma_{\rm max}},
thereby precipitating the exponential tails.  Given that \teq{n_e(\gamma )} is the 
distribution of electrons differential in Lorentz factor, this \teq{\gamma^2 n_e(\gamma )} 
representation highlights the fact that most of the synchrotron and SSC radiative power 
is generated by electrons with the Lorentz factors near and above \teq{\gamma_{\rm br}}.
The Mrk 501 case (blue histogram) had \teq{\eta_1=100} and \teq{\alpha = 1.5},
while the BL Lac one (red) had \teq{\eta_1=20} and \teq{\alpha = 3}.  The comparison 
\teq{n_e(\gamma )\propto \gamma^{-1}} differential distribution exhibited in green 
is the expected power law for high \teq{\eta_1} cases (see Section~\ref{sec:shock_accel}).
\label{fig:blazar_edist}}
\end{figure}

The electron distribution corresponding to the spectrum in
Fig.~\ref{fig:MW_spec_Mrk501} is illustrated in
Figure~\ref{fig:blazar_edist}, in a manner analogous to the \teq{\nu
F_{\nu}} representation: \teq{\gamma^2 n_e(\gamma )} distributions give
an approximate scaling of the emission power of synchrotron and SSC
signals from electrons of a given Lorentz factor \teq{\gamma}.  The
thermal Maxwellians are fairly prominent, yet they couple efficiently
and smoothly into the non-thermal accelerated population.  The
exponential turnovers near the maximum energies are precipitated by
efficient radiative cooling in the compact acceleration zone, depicted
in Fig.~\ref{fig:model_geom}.  This shuts off the acceleration, and the
electron distribution evolves through continued cooling in the larger
radiation zone. Cooling ceases when the radiation cooling length becomes
comparable to the size of the radiation region, which is specified by
the parameter \teq{R_{\rm rad}} in Table 1, and this introduces the
break at \teq{\gamma\sim \gamma_{\rm br}}, the approximate value of
which satisfies Eq.~(\ref{eq:gamma_br_cool}). For the extended state of
Mrk 501, since the synchrotron component slightly dominates the SSC one
in terms of the energy flux, the inverse Compton mechanism contributes
the minority of the cooling; for any considerations of the high state,
the situation would be reversed.  For each process, efficient cooling
steepens the electron distribution by an index of unity above the
cooling break, and this induces the well-known steepening by an index of
\teq{1/2} in the photon spectrum.  Such a break is seen at around
\teq{10^{12.5}}Hz for the synchrotron component in
Fig.~\ref{fig:MW_spec_Mrk501}, and is barely discernible at around
\teq{10^{19}}Hz for the SSC contribution therein.

We note in passing that the injected shock acceleration distributions
that led to those depicted for the larger cooling zone in this Figure
have most of their energy allocated to the particles near \teq{\gammax}.
 In such cases, the acceleration simulations need to be upgraded to
account for non-linear modifications to the shock MHD structure imposed
by the pressure of the energetic non-thermal particles.  This feedback
phenomenon is well understood in non-relativistic shock theory
\citep[e.g.][]{EE84,EBJ96,BERGG99}, and motivates future refinements to
blazar studies of the genre herein, using extensions of this feedback
theory to relativistic shocks along the lines of \citet{EWB13}.

Beholding the multiwavelength spectrum, at first sight, the character of
the Mrk 501 SSC model fits here resemble those in numerous expositions
on this source, including \citet{Acciari11} and \citet{Furniss15}. 
However there is a key development here beyond phenomenological electron
distributions, namely broken power laws that are truncated at a minimum
electron Lorentz factor, which are usually invoked in other studies of
Mrk 501 \citep[e.g., see][]{Aleksic15} and various other blazars.  These
forms are unphysical from the standpoint of shock acceleration theory. 
Here, {\it we have complete thermal-plus-non-thermal distributions at
our disposal}, and this serves to advantageously define the thermal
plasma density normalization relative to that of the accelerated
electron population (see Fig.~\ref{fig:blazar_edist}).  The plasma
density connects to determination of the shock structure, and the number
density of accelerated electrons benchmarks the radiative output and
also the jet kinetic energy \teq{L_{\rm kin}} via Eq.~\ref{eq:Lkin}.  In
addition, knowing the total number density \teq{n_e} fairly precisely
permits the determination of the electron plasma frequency
\teq{\omega_p}, and the plasma magnetization parameter \teq{\sigma =
\omegaB^2/\omega_p^2} (not possible for phenomenological power-law
distributions).  Inspection of Table~1 reveals that the magnetization is
\teq{\sigma\sim 1} for our Mrk 501 fitting, a result that will be
interpreted in Section~\ref{sec:discussion} as implying similar speeds
of acceleration for reconnection and for the diffusive Fermi mechanism
in the Bohm limit.  Also listed in Table 1 are determinations of the
hydrogenic Alfv\'en speed, in all cases virtually non-relativistic and
substantially inferior to the shock speed. These rise to relativistic
speeds with the addition of a dominant pair component to the jet.
Accordingly, here we forge an interconnection between photon emission
and underlying shock plasma properties in blazar jets, in a substantial
advance beyond previous works.

From the spectroscopic point of view, it is apparent from
Fig.~\ref{fig:MW_spec_Mrk501} that the thermal portion of the electron
distribution functions played no role in constraining parameter space.  
 The mildly-relativistic electrons in their thermal component are
inefficient generators of both synchrotron and SSC radiation. Thermal
synchrotron would appear in low frequency radio windows and is heavily
self-absorbed by the inverse synchrotron process, which is treated in
our radiation codes: the signatures of such attenuation generally appear
at frequencies below \teq{10^{10}}Hz when \teq{B\lesssim 10}Gauss.
Thermal SSC appears in the optical-X-ray window, and is swamped by the
strong X-ray synchrotron emission for Mrk 501.  Both contributions are
off-scale in the Figure.   Hence, in the case of Mrk 501, spectroscopy
associated with the thermal jet electrons is irrelevant.  This contrasts
the interesting case of AO 0235+164, our final object for study.

\subsection{Case Study: BL Lacertae}

To offer a picture distinct from that of Mrk 501, the focus turns to BL
Lacertae, the prototype BL Lac object, which has a much softer
synchrotron component that peaks in the optical (i.e., an LBL) and
dominates the hard X-ray and gamma-ray signals. It is also a nearby
source, at \teq{z=0.069}, and various multiwavelength campaigns have
been staged to help elucidate its character.  A survey undertaken in the
early days of the {\it Fermi} mission is the focal point here.  This was
for the extended period of August -- October 2008, and while the source
was variable and possessed episodes of relatively enhanced radiative
output, its broadband emission was not that of a characteristically high
state.

\begin{figure}
\vspace*{-10pt}
\centerline{\hskip 10pt\includegraphics[width=9.5cm]{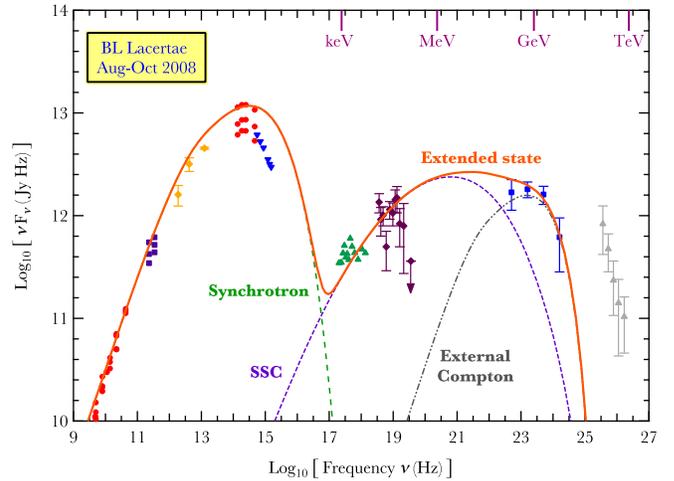}}
\vspace*{-5pt}
\caption{Multiwavelength \teq{\nu F_{\nu}} spectra (points) spanning the
radio, optical, X-ray and gamma-ray bands, together with model fits as
described in the text, for the August--October 2008 extended {\it
Fermi}-LAT observation of the blazar BL Lacertae. The data are taken
from the LBAS study of Abdo et al. (2010), and are briefly detailed in
the text: the colored data are approximately contemporaneous with the
{\it Fermi}-LAT collection.  The grey MAGIC points above \teq{10^{25}}Hz
are archival, being shown to illustrate the spectral variability between
different observational epochs; they were not accommodated in the model
fit. The broadband models consist of a synchrotron component (dashed
green curve) up to the optical band, an SSC component in the X-rays and
gamma-rays (dashed purple curve), and an external inverse Compton
emission component (dash-dot grey) seeded by IR photons from a dusty
accretion torus. The full jet model spectrum (orange) includes a
correction for \teq{\gamma-\gamma} absorption by the extragalactic
background light, and combined with the data, constrain the model
diffusion parameters to \teq{\eta_1=20} and \teq{\alpha =3} (see text).
\label{fig:MW_spec_BLLac}}
\end{figure}

The data compilation for this three month epoch is detailed in the {\it
Fermi}-LAT collaboration paper on a multitude of so-called LBAS blazars
\citep{Abdo10c}. The purpose of this compendium was a first quick-look
survey of the general character of bright blazar spectra in the {\it
Fermi} era.  Therefore, for BL Lacertae, the mutliwavelength data
presented in that paper are depicted in Fig.~\ref{fig:MW_spec_BLLac},
serving as our modeling benchmark. The data in color are from {\it
Fermi}-LAT in the 100 MeV to 30 GeV energy window (blue), Swift XRT
(0.3-10 keV, green) and Swift BAT (15-200 keV, purple) in the X-rays,
Swift-UVOT in the near ultra-violet, radio measurements from Effelsberg,
Owens Valley and RATAN, and data from other optical, infra-red and radio
facilities. Since not all were precisely contemporaneous with {\it
Fermi}-LAT observations and each other, modest mismatches between model
fits and data should not be over-interpreted. Synchrotron emission from
BL Lac is bright enough to outshine its host galaxy, and so no galaxy
contribution is discernible in the optical, unlike Mrk 501.
Fig.~\ref{fig:MW_spec_BLLac} also presents archival MAGIC data
\citep{Abdo10c} in grey that were not used in the fitting protocol, and
serve as a guide for the variable character of gamma-ray signals from BL
Lac.  Fig.~23 of \citet{Abdo10c} can be consulted for a more extensive
presentation of archival spectroscopy for BL Lac.

The radiation modeling protocols were implemented as described above,
using a similar suite of particle distributions generated by the Monte
Carlo shock acceleration simulation.  Radiation model and simulation
input parameters, and also those derived from the input ones, are listed
in Table~1.  The shock MHD parameters were identical to the Mrk 501
study. The turbulence parameters \teq{\alpha} and \teq{\eta_1} for the
simulations were again adjusted to optimize the fit, for which spectral
results are presented in Fig.~\ref{fig:MW_spec_BLLac}. The
radio-to-optical data dictate that the synchrotron component is
dominant, and turns over below the X-ray band.  This imposes a higher
value of the magnetic field and its associated portion of the total
energy budget than for the case of Mrk 501.  Concomitantly, synchrotron
cooling of electrons in the acceleration zone is more rampant, thereby
reducing the maximum Lorentz factor to \teq{\gammax\sim 3.5\times 10^3}
(see Fig.~\ref{fig:blazar_edist}). Accordingly, the synchrotron
self-Compton \teq{\nu F_{\nu}} peak energy appears in the soft
gamma-rays, below 10 MeV, as is evident in Fig.~\ref{fig:MW_spec_BLLac}.
This circumstance is essentially unavoidable in LBLs that are
synchrotron-dominated. Hence, the {\it Fermi}-LAT data require another
component to be present.

To describe such, here we employ an external Compton signal seeded by
emission in the near infra-red (NIR) from a dusty torus associated with
the accretion flow onto the central black hole.  These constitute an
isotropic thermal photon field with the luminosity and temperature as
listed in Table~1. It is a simple matter to discern why this EC
component emerges in the {\it Fermi}-LAT band.  While the torus appears
in the NIR for an observer, as an external isotropic field proximate to
the jet, it is boosted in the jet frame into the far-UV. This is at
considerably higher frequency than the synchrotron emission, which
clearly peaks in the IR band in the jet frame.  Accordingly, the EC peak
is substantially bluer than that of the SSC in both the jet frame, and
our observer frame, by over a factor of 100 higher in energy. Moreover,
the Doppler-boosted flux enhancement of the torus seed photons in the
jet frame can generate a gamma-ray \teq{\nu F_{\nu}} flux comparable to
the SSC without the seed NIR signal being discernible above the very
strong, beamed, synchrotron component at those frequencies.  The breadth
of the EC peak is restricted by the intrinsic width of the Planck
spectrum from the torus.

While this introduces an extra subset of model parameters, there appears
no likely alternative in one-zone leptonic models.  Similar broad
spectral structure from X-rays to hard gamma-rays could be generated by
a multi-zone blazar emission model: such was the protocol adopted by
\citet{ZW16} in the modeling of AP Librae, an LBL object whose
multiwavelength spectroscopy is very similar to that of BL Lacertae. A
possible discriminator between these two competing pictures is the
appearance of a flat bridge in the spectrum from our protocol between
the two inverse Compton components.  Such a feature might not appear in
a multi-zone construction, which may evince broader spectral curvature.
This provides motivation for future sensitive gamma-ray detectors below
the {\it Fermi}-LAT window, perhaps employing Compton telescope
technology.

At the very highest energy gamma-rays, as for Mrk 501, our modeling
incorporated a \teq{\gamma\gamma\to e^+e^-} pair opacity correction
\citep{FRD10} due to the intervening EBL.  However, since the EC peak in
BL Lac lies at 10 GeV and well inside the LAT window, this correction is
quite small, and did not influence our fitting protocol.  It is
anticipated that for this source, when it is in a hard state as captured
in the non-contemporaneous MAGIC data, such attenuation compensations
will come into play.

It is notable that BL Lacertae is singular in our ensemble of blazars in
that it has measurements of the jet bulk velocity.  \citet{Jorstad05}
outline VLBI imaging, polarimetry and variability studies of a number of
radio jet sources, including BL Lac.  Combining imaging and flux
variability information at 7mm, they were able to isolate both the
apparent superluminal speed of the jet and the Doppler factor. This
yielded a determination of the bulk Lorentz factor of \teq{\Gamma_{\rm
jet}\sim 7\pm 2} for BL Lac: see Table~11 of \citet{Jorstad05} and
associated discussion. In a later investigation of TeV flaring in BL Lac
using VERITAS, \citet{Arlen13} employed internal \teq{\gamma\gamma\to
e^+e^-} pair transparency arguments based on \citet{DG95} to constrain
the jet bulk motion to Lorentz factors \teq{\Gamma_{\rm jet}\gtrsim 
13-17}.  \citet{Arlen13} argued that this determination was possibly not
in conflict with the lower value obtained in the VLBI study, with the
gamma-ray emitting region being distinct from the radio ones and perhaps
lying closer to the central black hole, thereby invoking a decelerating
jet picture.  Since these gamma-ray constraints are most germane to our
multiwavelength investigation here, they guided our adoption of
\teq{\Gamma_{\rm jet}=15} for the models of BL Lac.  This guaranteed
pair transparency for the jet for gamma-rays below 1 TeV.  Higher values
of \teq{\Gamma_{\rm jet}} that permit jet pair transparency can also
lead to viable fits with the same particle distributions from
acceleration theory, yet require reasonable adjustments of environmental
parameters such as the field strength and electron number density.

As with Mrk 501, the upshot of the M/W modeling is that a large value of
\teq{\alpha} is required to generate relatively efficient injection of
electrons from thermal energies into the acceleration process, and
inefficient acceleration at the highest Lorentz factors.  Notably, since
BL Lac is an LBL blazar, a value of \teq{\alpha \sim 3} was required to
increase \teq{\eta (\gammax )} above \teq{10^7} and move the synchrotron
peak into the optical band. The values we obtained, \teq{\eta_1=20} and
\teq{\alpha =3} (i.e. \teq{\eta (p) \propto p^2}), indicate diffusion
well away from the Bohm limit at all momenta, and that the level of
turbulence must decline with distance from the shock even more
profoundly than for Mrk 501. As with our first case study, there is a
tolerance of about \teq{\pm 0.2} in the range of \teq{\alpha}, and a
tolerance of a factor of \teq{\sim 1-5} in \teq{\eta_1} permitting M/W
spectral fits of similar quality.  The value of \teq{\eta_1=20} obtained
for BL Lac is noticeably lower than for our two other case study
blazars.  Yet, it is also somewhat higher than the value of
\teq{\eta_1\sim 5} determined from fitting thermal+non-thermal
distributions of protons and alpha particles that are measured {\it in
situ} at interplanetary shocks in the non-relativistic solar wind
(Baring et al. 1997). This suggests that mildly relativistic shocks in
blazar jets are less efficient at injecting charges into the
acceleration process than are traveling shocks in the solar wind. Note
that the contributions to each radiation component from the thermal
electron population are small enough to be off scale in the Figure. The
electron distribution corresponding to the fit is depicted in
Fig.~\ref{fig:blazar_edist}.

An issue that has not yet been addressed concerns the variability
timescale, which must be large enough to accommodate the time for
acceleration up to the largest particle energies required to fit the
spectral data. The acceleration rate is given by \teq{d\gamma /dt\sim
\omegaB/\eta} in Eq.~(\ref{eq:acc_rate}), assuming a relativistic shock.
Thus, the most energetic charges are accelerated on timescales of 
\begin{equation}
   \tau_{\rm acc} \;\sim\; \dover{\eta (\gamma_{\rm max}) \, \gamma_{\rm max}}{\omegaB}
 \label{eq:}
\end{equation}
in the rest frame of the jet.  Each of these quantities on the right
hand side is listed in Table 1, yielding derived values of the
characteristic cumulative timescale for acceleration \teq{\tau_{\rm
acc}}, also listed in the Table.  Thus we infer \teq{\tau_{\rm acc} \sim
4\times 10^5}sec for Mrk 501, i.e. around 4.5 days. Similarly,
\teq{\tau_{\rm acc} \sim 2\times 10^4}sec for Bl Lac. To connect to the
observer's frame, these are shortened by time dilation factors of
\teq{1/\Gamma_{\rm jet}} for transverse variability, i.e to around
\teq{1.6\times 10^4}sec or 4 hours for Mrk 501, and around 20 minutes
for Bl Lac.  Such dilation applies to fluctuations in the direction
normal to the line of sight to the observer, i.e. approximately the jet
axis.  Longitudinal variability along the jet, such as are expected from
colliding shells, reduces the acceleration timescales by another factor
of \teq{1/\Gamma_{\rm jet}}, rendering them very short in the observer
frame.  These are then less than the variability timescales for both
sources in the hard gamma-rays (e.g., see Aliu et al., 2016, for Mrk 501
data in the high state that was excluded from the epoch considered in
this paper), for both extended states and flaring episodes within the
observing epochs under consideration, and so no internal inconsistency
arises.

\subsection{Case Study: AO 0235+164}

To provide another contrasting case, we turn to the LBL blazar AO
0235+164, which has a radio-optical component that turns over below the
UV band.  It is normally weak in X-rays, however, in flare state, it
possesses a significant X-ray excess.  Such enhancements have been
discussed \citep{SMMP97,BSMM00} as a possible signature of a bulk
Comptonization effect \citep[see also][]{SBR94}, where hot thermal jet
electrons upscatter an ambient, quasi-thermal radiation field that
manifests itself in the infra-red, optical or UV bands. This seed field
could be optical/UV radiation from the broad-line or narrow-line
regions, or IR emission from warm dust in the greater AGN environment;
it is distinct from the non-thermal jet synchrotron emission.  One of
the objectives here is to explore whether the X-ray ``excess'' can be
attributed to the substantial thermal electron pool evinced in shock
acceleration distributions such as those depicted in
Fig.~\ref{fig:accel_dist}.  In the fast-moving jet, this population may
produce significant radiative signatures due to Comptonization of an
external radiation field, taking advantage of the same Doppler boosting
enhancement that was captured in our modeling of BL Lac.

\begin{figure}
\vspace*{-10pt}
\centerline{\hskip 10pt\includegraphics[width=9.5cm]{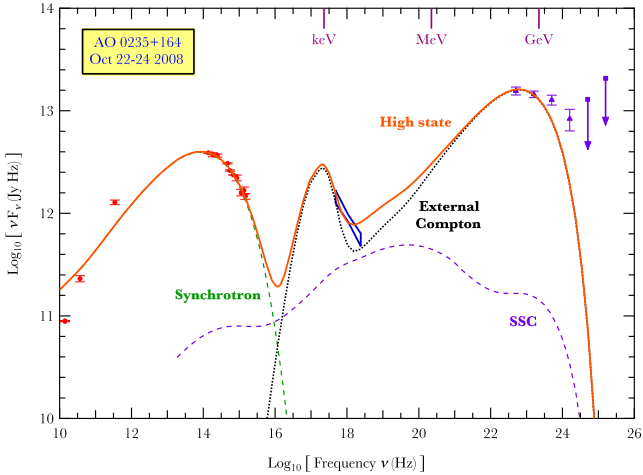}}
\vspace*{-5pt}
\caption{Multiwavelength \teq{\nu F_{\nu}} spectra (points) spanning the
radio, optical, X-ray and gamma-ray bands, together with model fits as
described in the text, for the October 2008 high state {\it Fermi}-LAT
observation of the blazar AO 0235+164. The campaign data are taken from
\citet{Ackermann12}, corresponding to the high state epoch during
October 22-24.  The gamma-ray detections and upper limits are from {\it
Fermi}-LAT, while the X-ray ``butterfly'' block (blue) represents {\it
Swift} XRT data. The broadband models consist of a synchrotron component
(dashed green curve) up to the optical band, a two-order SSC
contribution in the optical, X-rays and gamma-rays (dashed purple
curve), and bulk Comptonization emission (EC: dotted black curve)
between \teq{\sim 0.1}keV and \teq{\sim 10}GeV. The total model spectrum
(orange) includes a very small correction for \teq{\gamma\gamma}
absorption by the extragalactic background light, and combined with the
data, constrains the model diffusion parameters to \teq{\eta_1=225} and
\teq{\alpha =3} (see text).
\label{fig:MW_spec_AO0235}}
\end{figure}

The focus here is on modeling data collected during the flaring episode
of September -- November 2008.  In particular, we employ the
multi-wavelength campaign observations discussed in \citet{Ackermann12};
see also \citet{Agudo11} for additional flare light curves and
polarization data.  The broadband spectrum for the high state portion of
the outburst, during the interval MJD 54761-54763 (October 22-24, 2008),
is depicted in Fig.~\ref{fig:MW_spec_AO0235}.  This appears in Fig.~7 of
\citet{Ackermann12}, wherein it can be compared with a low state
spectrum obtained about six weeks later that does not contain a {\it
Swift} XRT detection.  The list of observatories and missions
participating in this campaign is extensive --- it spans radio and
optical, with {\it Swift} XRT and RXTE providing X-ray monitoring, and
{\it Fermi}-LAT providing the principal gamma-ray observations.  The LAT
spectrum displays a break or possible turnover at energies around
\teq{3-4}GeV, with a power-law index of about \teq{2.1} below this
feature; the source is not detected in the TeV band \citep{Errando11}. 

The major model components and total multi-wavelength spectrum in our
fit of the flare data are also exhibited in
Fig.~\ref{fig:MW_spec_AO0235}. The jet and disk parameters used to
derive the model fit are listed in Table 1, and we remark that they are
somewhat different from those in our earlier introductory exposition
\citep{BBS14} on this source. Notably, the magnetic field at
\teq{2.5}Gauss was sub-equipartition by about a factor of 16.  To place
the synchrotron turnover frequency in the optical band (\teq{\sim
10^{14}}Hz), the values of \teq{\eta (\pmax )} and \teq{B} have been
chosen so that the maximum electron Lorentz factor in the comoving frame
of the jet is \teq{\gammax \approx 1.6\times 10^3}.  The shock obliquity
was modestly higher than for the BL Lac and Mrk 501 studies, a choice
that renders the thermal electrons comparatively more numerous. As with
BL Lac, the model fit required \teq{\alpha =3} and \teq{\eta_1=225}, so
that the diffusive scales of energetic particles are very large.  The
tolerances in these parameters that permit M/W fits of similar calibre
are about \teq{\pm 0.2} in the range of \teq{\alpha}, and a factor of
\teq{\sim 1-3} in \teq{\eta_1}. The moderate value of \teq{\gammax} then
positions the peak SSC frequency at around \teq{\gammax^2
10^{14}\hbox{Hz}\sim 6\times 10^{20}}Hz, i.e. around 5 MeV, as is
evident in Fig.~\ref{fig:MW_spec_AO0235}.  This is not the maximum
extension of SSC component: a second-order IC image of the synchrotron
continuum appears up to about 10 GeV.  A cooling break is also apparent
in the SSC spectrum at around a few keV.   Observe that the SSC
component is of insufficient luminosity to model the {\it Swift} XRT and
{\it Fermi}-LAT signals --- another component is needed to explain them.

These two high state detections are modeled here using
bulk-Comptonization/inverse Compton scattering of an external radiation
field.  This field is presumed to be quasi-isotropic in the observer's
frame, so that the jet material moves at \teq{\Gamma_{\rm jet}=35}
relative to these seed photons, and so can inverse Compton scatter them,
i.e. increase their frequency by a factor of \teq{\sim 4 \Gamma_{\rm
jet}^2}.  This choice of \teq{\Gamma_{\rm jet}} lies between that
adopted in some of the supporting modeling in the {\it Fermi}-LAT paper
of  \citet{Ackermann12}, namely \teq{\Gamma_{\rm jet}=20}, and the much
higher value of \teq{\Gamma_{\rm jet}\sim 50} indicated by structure
variability in the VLBA imaging over extended epochs that was presented
in \citet{Jorstad01}. We note that these values exceed the typical range
of \teq{\Gamma\sim 10-20} quoted in \citet{Ghisellini14} for BL Lac
objects where Compton dominance (or not) in SSC models is used to
constrain \teq{\Gamma} --- clearly, employing external Compton scenarios
will modify these estimates. The Lorentz factor of the thermal electrons
in the distributions of Fig.~\ref{fig:accel_dist} is less than about 2,
thereby modifying this IC energy enhancement by at most a factor of a
few.  Hence, to explain the {\it Swift} XRT flaring flux, the seed
background radiation needed a temperature of \teq{T\approx 10^3}K. This
corresponds to infra-red radiation, probably from a dusty torus.  The
thermal portion of the electron population boosts the ambient photons up
to the X-ray range, and roughly produces the steep {\it Swift} XRT
spectrum if it corresponds to the higher momentum side of the
Maxwell-Boltzmann distribution, as is clearly evident in
Fig.~\ref{fig:MW_spec_AO0235}.  This then connects to a very flat
external Compton tail, generated by the broken power-law tail portions
of distributions like those depicted in Fig.~\ref{fig:accel_dist},
modulo the cooling breaks discussed above. The IC  tail possesses a flat
spectral index of around 3/2, rising in the \teq{\nu F_{\nu}}
representation to meet the {\it Fermi}-LAT flux above 100 MeV.  The
large IC flux indicates very strong Compton cooling of electrons, and
this generates a very low cooling break energy at \teq{\gamma_{\rm
br}\sim 2}.

Note that despite the high redshift (\teq{z=0.94}) of this source
relative to those of the other two blazars, the EBL absorption
correction is minimal since the spectrum extends only up to around 20
GeV, i.e., similar to the situation for BL Lac. A peculiarity of this
blazar is that sensitive optical/UV spectroscopy reveals the presence of
an intervening damped Lyman-$\alpha$ absorbing system at a redshift of
\teq{z=0.524} \citep[e.g.,][and references therein]{Junkk04}. This
absorber manifests itself via the appearance of a broad, redshifted
2175\AA\ feature that is attributed to the presence of graphitic dust
grains nearer than AO 0235+164.  Such a concentrated repository of dust
(perhaps in a galaxy) along the line of sight can also provide
\teq{\gamma\gamma} pair absorption for the AO 0235+164 spectrum.  Yet
this source of opacity would again be confined to the energy band
greater than 20 GeV, and so has minimal impact on the spectral modeling
here.  This system can, however, serve to attenuate soft X-rays via
photoelectric absorption.

In modeling the broadband data, the ephemeral {\it Swift} X-ray
component is considered to be a bulk Comptonization signal, which
typically span a narrow waveband. \citet{SMMP97} demonstrate that such a
component needs to be generated at perhaps 100 gravitational radii or
more in order for the kinetic luminosity of the jet to not overproduce
X-rays.  The maximum mean free paths \teq{\lambda_{\parallel}(\gammax)}
listed in Table~1 are consistent with this constraint in systems where
the central black hole exceeds around \teq{10^8M_{\odot}} in mass.
\citet{BSMM00} note that such Comptonization X-ray signals are rare in
the OVV  quasar population. AO 0235+164 is peculiar in exhibiting such a
steeply declining X-ray spectrum, contrasting BL Lac as depicted in
Fig.~\ref{fig:MW_spec_BLLac}. It is natural to ask whether the X-ray
flare in AO 0235+164 could alternatively be a distinct synchrotron
component from a different region in a multi-zone construction.  This
requires more model parameters, with no additional constraints. To
render this explanation viable, either the magnetic field of the second
zone should be much larger, or the maximum Lorentz factor \teq{\gammax}
should be much higher, or both.  Either of these requirements would also
demand a smaller cooling volume so as to reduce the X-ray luminosity
below the optical one.  If \teq{\gammax} is much larger than the value
of \teq{1.6\times 10^3} listed in Table 1, one anticipates the
appearance of a significant SSC component extending up into the TeV
band. At MeV energies, this will necessarily lie below the one displayed
in Fig.~\ref{fig:MW_spec_AO0235} that corresponds to a synchrotron fit
in the radio to optical.  Accordingly its extension out to the LAT band
would likely lie below the observational data, and one envisages
difficulties for a picture with two distinct synchrotron (and also SSC)
components without an external Compton one.  While it is possible, in
principle, to construct a viable two-zone synchrotron plus SSC plus EC
model for the AO 0235+164 multiwavelength spectrum, we view the
interpretation of the steep X-ray continuum in the high state as a bulk
Comptonization signature as the simpler explanation.

\vspace{-10pt}
\section[]{Discussion: Turbulence \& Diffusion}
 \label{sec:discussion}

The recurring theme in all of these multi-wavelength spectral fits is that 
the diffusion of charges in blazar acceleration zones should be well removed 
from the Bohm limit \teq{\lambda \sim r_g} at nearly all energies, and furthermore
that the diffusion scale \teq{\lambda} should lengthen fairly rapidly with increasing 
particle momentum.  This indicates that the physical environment of the
blazar shocks should possess a diminished level of turbulence at larger
distances from the discontinuity, precipitating the longer diffusive mean free
paths for larger momenta.  This circumstance is not entirely without
parallel.  In the heliosphere, the solar wind contains traveling
interplanetary (IP) shocks, at which accelerated ions are clearly detected, 
and MHD turbulence is measured using magnetometers mounted
on {\it in situ} spacecraft. 
The turbulence near such IP shocks is observed to satisfy \teq{\langle\delta
B/B\rangle \sim 0.1}, and the diffusive transport of suprathermal H and He ions 
at \teq{\sim 10} keV energies is
inferred to be not very different from the Bohm limit, with mean free paths 
of \teq{\lambda_{\parallel} \lesssim 10^4}km \teq{\sim 10^{-4}}AU \citep[e.g.,][]{BOEF97}.  In
contrast, away from these shocks, the turbulence is much more benign, and
the diffusive mean free paths \teq{\lambda_{\parallel}} of 10 MeV--1 GeV ions are much larger 
\citep{FJO74,Palmer82,Bieber94,ZMBM98}, on the scale of \teq{0.1-3}AU or so.

\subsection{Turbulence, Diffusion and the Solar Wind}

From such contrasting information, \teq{\alpha >1} circumstances must clearly exist.  
Determinations of the diffusion index in the solar wind are non-trivial, and 
not all that precise.  The reason for this is primarily that 
magnetometer measurements consist of three-dimensional vector fields
{\bf B} at one particular space location.  This restricts knowledge of the
global field character, and prohibits assessment of true diffusive trajectories of 
charges on extended spatial scales.  For examples of ion diffusion characteristics 
inferred from older solar wind data, the reader can consult \citet{EMP90}
for observations at the Earth's bow shock, which indicate that \teq{1/2
< \alpha < 3/2}.  In contrast, lower values \teq{\alpha\sim 1/2} are 
obtained from turbulence in the interplanetary magnetic field
on larger scales at heliodistances around 20 AU \citep{Moussas92}, 
and in addition \teq{1/2 < \alpha < 4/5} can be deduced from ions accelerated in solar particle
events \citep{MGH83} close to the sun.  It is immediately obvious
that these values of \teq{\alpha} do not match the larger values inferred
from the blazar spectral modeling.  Accordingly, different conditions must apply 
to the energetic particles in the blazar jet environment, as will become apparent shortly.

\begin{figure}
\vspace*{-10pt}
\centerline{\hskip 10pt\includegraphics[width=9.3cm]{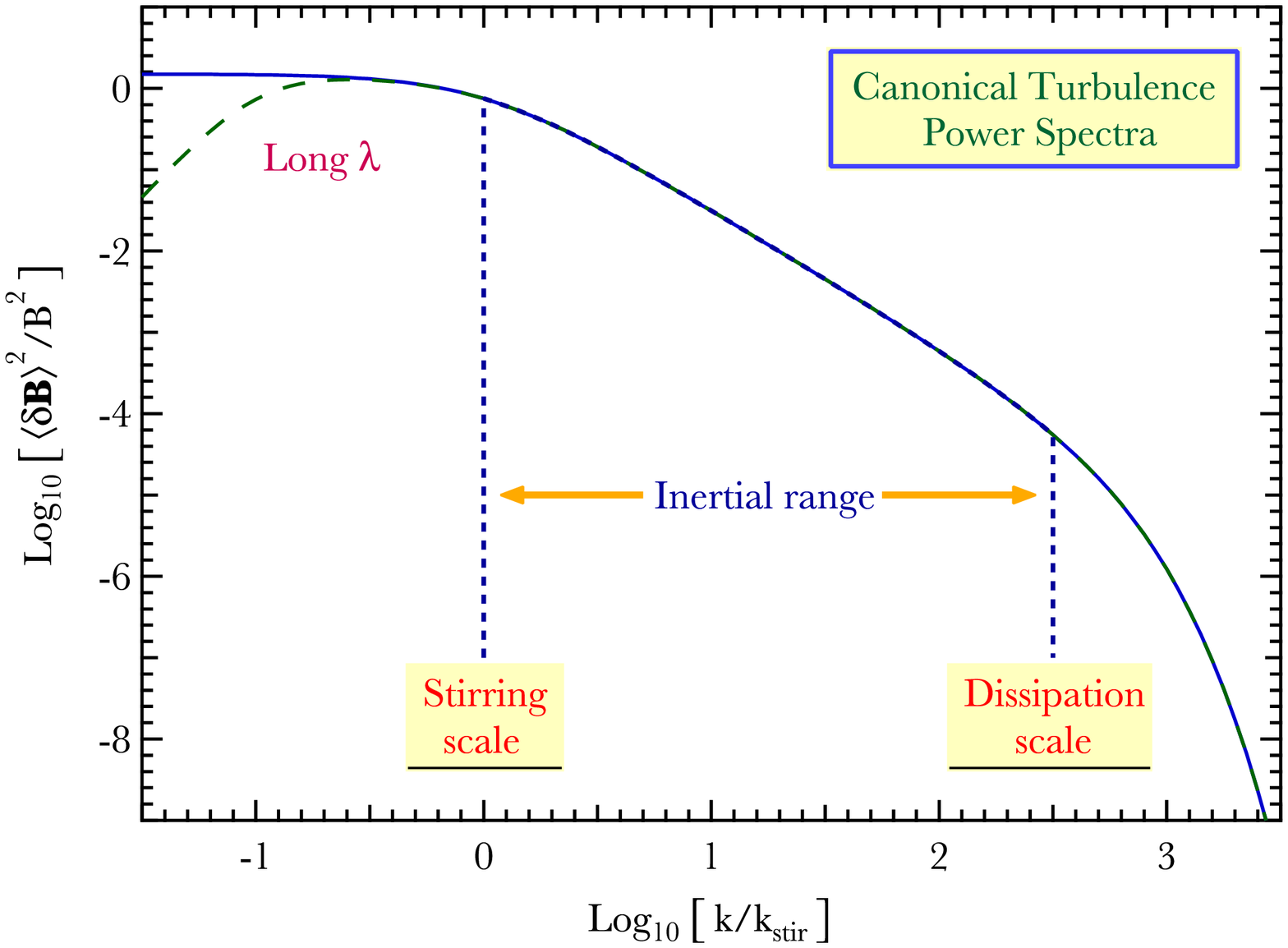}}
\vspace*{-5pt}
\caption{Schematic plot of a representative power spectrum 
\teq{\langle \delta \mathbf{B}\rangle^2/B^2} for 
cascading MHD turbulence in astrophysical systems like the solar wind, 
interpreted as approximately what should exist in blazar jets.
The turbulence is seeded at low wavenumbers {\bf k} (long wavelengths)
around a stirring scale of \teq{\vert \mathbf{k}\vert\sim k_{\rm stir}} and 
cascades in a roughly scale-independent manner to high frequencies 
or short wavelengths.  Eventually it reaches a dissipation scale, 
here chosen to be \teq{\vert \mathbf{k}\vert \sim 10^{2.5}k_{\rm stir}},
where absorption of MHD waves by the plasma heats the charges. 
The power spectrum in the intermediate wavenumber domain, the 
inertial range that is usually what is captured in solar wind magnetometer 
data, is shown here for a Kolmogorov scaling 
\teq{\langle \delta \mathbf{B}\rangle^2/B^2\propto k^{-5/3}} in one-dimensional MHD turbulence
(see text); other approximate scalings are realized in Nature.  At extremely long 
wavelengths \teq{\lambda \gg 2\pi /k_{\rm stir}} beyond the stirring scale, 
the precise form of the power spectrum is usually unknown,
and is represented by the two cases shown, namely flat and declining.
\label{fig:powerspec}}
\end{figure}

More recent magnetometer measurements of the quiet solar wind \citep{PRG07} 
by the {\it Wind} spacecraft over a period of several years captures a broader
dynamic range for the power spectrum.  The {\it Wind} data indicate that the
inertial range for turbulence \teq{\langle (\delta B)^2/8\pi \rangle}
spans 3-4 decades in frequency above \teq{\nu_{\rm stir}\sim 3\times
10^{-5}}Hz, which marks the stirring scale of the system, i.e., 
\teq{v_{\rm sw}/\nu_{\rm stir} \sim 10^7}km, where \teq{v_{\rm sw}\sim 400}km/sec is the 
approximate speed of the solar wind.  The spectrum for this inertial range is fairly close to the standard
Kolmogorov \teq{\nu^{-5/3}} form.  Below this frequency, the turbulence
spectrum flattens, resembling the shape depicted in the schematic of 
Figure~\ref{fig:powerspec}, and one anticipates that diffusion in solar wind
plasma turbulence must decline.  The normalization of the power spectrum 
establishes the diffusive scale of ions in such turbulence, and estimates
using quasi-linear diffusion theory lead 
to the aforementioned diffusive scale of the order of \teq{0.1-3}AU in the undisturbed 
interplanetary medium.  Note that departures from pure Kolmogorov forms 
for solar wind field turbulence are observed in high heliographic latitude 
(i.e. well out of the ecliptic) {\it Ulysses} magnetometer determinations
\citep{Goldstein95}.  This excursion should alter the momentum dependence 
of the mean free path somewhat. 

An abundance of data in the heliosphere indicates that the solar wind is 
very structured, with active regions containing shocks, enhanced turbulence and the 
production of energetic ions and electrons.  Perhaps the popular picture 
that shocks generate turbulence through their viscous dissipation, 
and that this is muted or damped in regions remote from them, 
applies to both the solar wind outflow and also to blazar jets.  This is the 
interpretative message encapsulated in this paper.  We now explore 
some elements of diffusion theory in the solar wind, bearing in mind that while
commonalities between the structured, non-relativistic heliospheric wind and 
turbulent relativistic blazar jets are expected to exist, there should also be some inherent 
differences between turbulence in these two contrasting cosmic plasma environments.

Appraising the assembly of solar wind data in terms of the stochastic impact of 
magnetic turbulence on charge motion requires a theoretical construct for 
modeling diffusion.  The historical tool dating from the 1960s has been the quasi-linear theory (QLT)
of diffusion in magnetodynamic turbulence, i.e. that embedded in non-relativistic flows,
for which electric field fluctutations are small.  There are many expositions of such in the 
literature, and here we capture their essence through a compact set of equations.
The choice here of outlining elements of QLT in non-relativistic environs is motivated by the 
ability to efficaciously define certain simple and important characteristics, principally 
the momentum dependence of the diffusive mean free path, and how this connects 
to the power spectrum of turbulence.  It is anticipated that the global character 
of \teq{\lambda_{\parallel}(p)} being a strongly increasing function of momentum
will carry over into the mildly relativistic plasma regime that is germane to 
our blazar jet frame context.

In the diffusive limit of particle transport, the spatial development of 
particle distribution functions along the mean magnetic field scales with the 
spatial diffusion coefficient \teq{\kappa_{\parallel} = \langle (\Delta x)^2\rangle/(2\Delta t)}, 
which appears in the second-order derivative terms in the Fokker-Plank form
of the transport equation.  In the limit of transport subject to small, chaotic field 
fluctuations, charges diffuse in their pitch angle
\teq{\mu = \mathbf{v} \cdot \mathbf{B} / (\vert \mathbf{v}\vert . \vert \mathbf{B}\vert )},
so that \teq{\kappa_{\parallel}} can be related \citep[e.g.,][]{Jokipii66,KP69,HW70,Schlick89}
to the pitch-angle diffusion coefficient via
\begin{equation}
   \dover{\lambda_{\parallel}v}{3}\; =\; \kappa_{\parallel} 
   \;\equiv\; \dover{\langle (\Delta x)^2\rangle}{2\Delta t}
   \; =\; \dover{v^2}{8} \int_{-1}^1 \dover{(1-\mu^2 )^2}{D_{\mu\mu}}\, d\mu\quad ,
 \label{eq:diffusion_coeff}
\end{equation}
for charges of speed \teq{v}.  The relationship between \teq{\kappa_{\parallel}}, 
\teq{D_{\mu\mu}}  and the mean free path \teq{\lambda_{\parallel}} parallel 
to the mean field presumes isotropy of the diffusion process.  Using QLT, the 
pitch angle coefficient can be approximately related to the power spectrum 
of field turbulence:
\begin{equation}
   D_{\mu\mu} \;\equiv\; \dover{\langle (\Delta \mu)^2\rangle}{2\Delta t}
   \; \approx\; \dover{\pi v}{r_g^2B^2} \, \dover{1-\mu^2}{\vert \mu\vert}
   \, {\cal P} \left( k_{\parallel} = \dover{\omegaB}{\gamma v\,\vert \mu\vert} \right)\quad .
 \label{eq:Dmumu_QLT}
\end{equation}
In this expression, \teq{r_g =pc/(eB) = \gamma v/\omegaB} is the
gyroradius of the charge, as employed throughout this paper, and \teq{B}
is the strength of the mean unperturbed field. Note that slightly
different constructions for \teq{D_{\mu\mu}} appear in the literature. 
The function \teq{{\cal P}(k_{\parallel})} is the power spectrum of
one-dimensional ``{\it slab}'' turbulence with propagation wavenumbers
\teq{k_{\parallel}} along the mean field direction; it is essentially a
Fourier transform of the spatial fluctuations in \teq{(\delta B)^2}. 
The coupling between the power spectrum and the diffusion coefficient
implicitly presumes that the turbulence is propagating at
non-relativistic speeds, i.e., is magnetostatic, so that the applicable
diffusion tensor is three-dimensional.  Details of this derivation are
presented in \citet{Jokipii66,FJO74,Lee82,ZMBM98}, and numerous other
papers in the literature.

A key element of this result is that it presumes resonant interactions
between charges and waves at the Doppler-shifted cyclotron frequency,
i.e. \teq{k_{\parallel}v\, \vert \mu\vert = \omegaB/\gamma}. This is a
circumstance that is not always realized in plasma turbulence, as will
become apparent below. Departures from this idealized gyroresonant case
have been presented in \citet{Schlick89}, for example, wherein
embellishments such as incorporating the influence of finite Alfv\'en
wave speeds can ameliorate the problem of the inherent suppression of
diffusion through pitch angles \teq{\arccos \mu = \pi /2}.
Notwithstanding, the combination of Eqs.~(\ref{eq:diffusion_coeff})
and~(\ref{eq:Dmumu_QLT}) serves to define the generic character of
diffusion in turbulent fields.

To this purpose, one can follow Appendix A of \citet{ZMBM98} and
approximate the power spectrum of one-dimensional turbulence with a
modified Kolmogorov form:
\begin{equation}
   P(k) \; \propto\; \dover{1}{(1+k^2/k^2_{\rm stir})^{5/6}}\quad .
 \label{eq:power_spec_Kolmogorov}
\end{equation}
This possesses true Kolmogorov dependence \teq{P(k)\propto k^{-5/3}} 
in the inertial range where \teq{k\gg k_{\rm stir}}, and is constant at 
wavenumbers below the stirring or field coherence scale \teq{k_{\rm stir}}.  
Such a form resembles the solid curve in the schematic plot in 
Fig.~\ref{fig:powerspec}, specifically in the inertial range below the 
dissipation scale at small wavelengths.  Inserting it into the core 
QLT equations above yields 
\begin{equation}
   \lambda_{\parallel} \;\propto\; \dover{3B^2}{4\pi} \, r_g^2
   \int_0^1 \mu\, (1-\mu^2) 
   \left\{1 + \dover{1}{(r_g \mu k_{\rm stir})^2} \right\}^{5/6} \, d\mu\; . 
 \label{eq:mfp_QLT}
\end{equation}
This is essentially the form given in Eq.~(A4) of \citet{ZMBM98}.  For 
low momentum charges, those with \teq{r_g k_{\rm stir}\lesssim 1}, 
the inertial range is accessible, and the dominant contribution 
to the integral comes from small pitch angle cosines \teq{\mu}, 
so that \teq{\lambda_{\parallel}\propto r_g^{1/3} \propto p^{1/3}}, 
approximately.  This behavior, which is commensurate with the values 
discussed above for heliospheric shocks, is predicated on the presumption 
of 1D Kolmogorov turbulence, and adjusts according to changes in the 
power spectrum and dimensionality of the turbulence.

The conclusion of generally weak momentum dependence of the diffusion
mean free path at low momenta is not restricted to QLT considerations.
Complementary tools for modeling diffusion in MHD turbulence are plasma
simulations, of which there are numerous analyses in the heliospheric
literature.  Hybrid plasma simulations \citep{GBS92,GJ99} in which
electrons are treated as a background fluid, but ions are modeled
kinetically, suggest mean free paths of non-relativistic ions obeying
Eq.~(\ref{eq:mfp_alpha}) with \teq{\alpha \sim 2/3} in one-dimensional
Alfv\'en-like turbulence.  Similiar diffusion scales are realized in the
solar wind turbulent diffusion analysis of \citet{ZMBM98} that combines
slab and 2D turbulence. Therein, the actually momentum dependence is not
well-constrained, though it is roughly consistent with QLT predictions. 
More recently \citet{CS14} computed diffusion coefficients using hybrid
plasma simulations of non-relativistic  shocks with strong turbulence. 
These isolate ion diffusion that is fairly close to the Bohm limit
\citep[contrasting][]{ZMBM98}, with a momentum dependence
\teq{\kappa_{\parallel}\propto p^{\alpha +1}} over roughly two orders of
magnitude that is slightly weaker (namely \teq{\alpha \sim 0.8-1}) than
for the Bohm case, i.e., \teq{\kappa_{\parallel}\propto
\lambda_{\parallel}v \propto p^2}. While the forgoing exposition
pertains to non-relativistic systems, {\it a priori} one anticipates
that its general character should also apply to relativistic plasmas,
where electric field fluctuations are no longer small.

Returning to inferences from formal QLT, in stark contrast to this low
energy domain, for charges with high momenta and therefore large
gyroradii, \teq{r_g k_{\rm stir}\gtrsim 1}, the Doppler resonance
condition \teq{k_{\parallel}v\, \vert \mu\vert  = eB/(\gamma mc)} is not
satisfied by turbulence in the inertial range, only by low frequency
waves, which are represented by the flat portion of the solid curve in
Fig.~\ref{fig:powerspec}.  The result from Eq.~(\ref{eq:mfp_QLT}) is
then \teq{\lambda_{\parallel}\propto p^2}, i.e. \teq{\alpha\sim 2} and
\teq{\eta \propto p}, corresponding to a strong momentum dependence.
This nuance is well-known in the heliospheric physics community. The
diffusion of {\it energetic charges} then receives a substantial
contribution from non-gyroresonant collisions with inertial turbulence. 
Such is expected to be the situation for electrons with large gyroradii
being accelerated in blazar shocks.

\subsection{Turbulence and Diffusion: Blazar Contexts}

The broadband spectral modeling presented for the three blazars clearly
indicates that plasma turbulence does not persist at the same levels out
to very large distances from blazar shocks, but instead declines in
intensity.  This applies in the inertial range, but can also be more
pronounced at wavelengths longer than the stirring scale. Such is the
circumstance represented by the dashed curve in
Fig.~\ref{fig:powerspec}, which is similar to the power spectra
exhibited in the theoretical modeling of Alfv\'en-like and compressible,
non-relativistic MHD turbulence that is presented in \citet{CL03}. For
this situation, one can anticipate that \teq{\alpha} can effectively
rise above two. Mathematically, this domain could be described by a
power spectral form \teq{P(k) \propto (k_{\rm stir}/k +k^2/k^2_{\rm
stir})^{-5/6}}, which would yield \teq{\alpha > 2} in
Eq.~(\ref{eq:mfp_QLT}). In other words, \teq{\alpha >2} scenarios are
entirely realistic for blazar multi-wavelength models.  Low levels of
long wavelength turbulence can be enough to eventually turn around the
highest energy charges that travel upstream in shock drift reflection
cycles. Note that it is not necessary that such turbulence be generated
within the shock zone: some portion of it may in fact be convected into
the shock from the remote regions of the jet, somewhat akin to the
turbulence fed into interplanetary shocks in the solar wind.

The discussion so far has focused on Alfv\'enic turbulence in
non-relativistic magnetized plasmas.  It is natural to ask whether the
properties of relativistic turbulence are similar.  Studies of
relativistic MHD turbulence are both more limited in number and more
recent, with the non-relativistic domain being deeply investigated
because of its relevance to the heliosphere, cosmic ray studies, the
interstellar medium and supernova remnants.  The upshot of recent work
on relativistic Alfv\'en-like turbulence \citep[e.g.,
see][]{Cho05,BL09,RR13,ZMacF13} is that a consensus appears to be
emerging that the power spectrum resembles the schematic depicted in
Fig.~\ref{fig:powerspec}, and that the inertial range sustains a form
fairly close to the classic Kolmogorov spectrum.  There may be some
differences in the anisotropy of turbulence from the non-relativistic
regime, and this is naturally expected due to the wave speeds
approaching \teq{c}. In particular, the 3D wavenumber profiles of Weibel
instability-driven turbulence near steep shocks inherently differ from
those for Alfv\'en-like turbulence propagating remotely from shocks, and
so the two should yield different anisotropies to the diffusion relative
to the mean field direction. The details of particle diffusive transport
in relativistic turbulence has not yet been fully explored.  It is
anticipated that scale-free power laws \teq{\lambda_{\parallel}\propto
p^{\alpha}} may well be realized for some isolated ranges of momenta. 
The spatial diffusion tensor will indeed be anisotropic, yet the basic
conclusion in our study of much greater turbulence near shocks than
remote from them will not be sensitive to the microdetails of the tensor
characteristics, nor more complicated forms for the
\teq{\lambda_{\parallel}(p)} function.

Our indication that blazar multiwavelength modeling leads to inferences
of weak turbulence and long diffusive mean free paths for the most
energetic leptons is not a consequence of the specific shock parameters
chosen in our case studies.  Varying the shock speed, magnetic field
obliquity, temperature and compression ratio from the values in Table 1
will lead mostly to adjustments in \teq{\eta_1}, and thereby the
efficiency of injection from the thermal population into the
acceleration process.  The derived values of \teq{\alpha >1} will not be
impacted by varying these shock properties to any significant degree,
since generating synchrotron turnovers in the optical or X-rays imposes
such \teq{\lambda_{\parallel}/r_g\gg 1} values at \teq{\gammax}.

Note that envisaging magnetic reconnection as an alternative scenario
does not alter this circumstance. Reconnection is generally invoked to
provide an acceleration mechanism (with \teq{E>B}) that is faster than
the cyclotron frequency. While perhaps desirable when modeling the GeV
gamma-ray flares detected by the {\it Fermi}-LAT in the Crab nebula,
this is the opposite of what is needed for blazars.  One can gauge the
relative efficacy of reconnection versus gyroresonant acceleration using
derived model parameters listed in Table~1.  Since the trapping of
electrons in relativistic reconnection zones in the PIC simulation of
\citet{SS14} is effected by the merging of magnetic islands traveling at
nearly \teq{c}, and the effective trapping ``radius'' is the inertial
scale \teq{\sim c/\omega_p}, the acceleration rate for relativistic
reconnection is of the order of \teq{\omega_p}. This is to be compared
with the Bohm-limited gyroresonant acceleration rate \teq{\omegaB}.  For
non-relativistic cases, their ratio is just \teq{\sqrt{\sigma}}, where
\teq{\sigma} is the plasma magnetization parameter in
Eq.~(\ref{eq:sigma_param}), which is around unity for our Mrk 501 and AO
0235+164 fits, and somewhat larger than unity for BL Lac.  Such an
estimate is only slightly modified by the Lorentz factor
\teq{\Gamma_1\lesssim 2} for our mildly-relativistic shocks.  This tells
us that magnetic reconnection is at most only slightly less efficient at
accelerating charges in our blazar shocks than gyroresonant
acceleration, an inference that would be impossible without our
virtually unique ability to benchmark the electron number density
relative to the non-thermal population.

It is salient to comment upon potential diagnostics on field turbulence
in blazars that can be afforded by radio and optical polarimetry. 
Polarization data require sufficient count statistics that generally
sample only larger spatial regions, unless intense flaring provides
sufficient flux to probe smaller scales.  In the case of BL Lac,
\citet{Gaur14} report monitoring in the V band spanning around 19 months
from May 2008 until January 2010 that exhibits variable polarization
degrees in the range \teq{\sim 5-30}\%.  Similar, but slightly smaller
values were observed in the R Band by \citet{Marscher08} during late
2005.  Comparable polarization degrees have been discerned for other
blazars, indicating the presence of moderately coherent fields.  This
ties more directly to an average MHD structure than it does to the
micro-turbulence being considered here. The scales for the diffusion of
all charges simulated in this study, even the most energetic ones, and
also those of the jet turbulence, are too small to be captured via
polarimetry, being far inferior to the sizes of causally-connected
regions inferred from polarization variability estimates. No conflict
emerges between the physical conclusions here of small levels of
turbulence far from shocks, and the moderate field coherence deduced
from the optical polarimetry of blazars.

As a final remark, this picture of large diffusion mean free paths for
the most energetic electrons present in blazar emission regions is not
confined to our subset of case study sources, which were selected to
span a significant range of blazar character.  The conclusion of
\teq{\eta (\gammax ) \gg 1} should be widely applicable to blazars of
both the HBL and LBL synchrotron spectral varieties in their compact
central regions.  It is notable that a similar conclusion of such large
\teq{\eta (\gammax )}, well above the Bohm regime, is made much further
out for the radio quasar 4C 74.26.  \citet{Araudo15} report that MERLIN
radio interferometer imaging in the proximity of the southern radio lobe
of this source has been able to map the bow shock zone and discern a
very restricted spatial scale. The narrowness of the emission zone is
far inferior to the radiation-reaction limited synchrotron cooling
length one infers for the shock zone by equating estimates like
Eqs.~(\ref{eq:cool_rate}) and~(\ref{eq:acc_rate}). This leads
\citet{Araudo15} to conclude both that wave damping must inhibit the
spatial scale of amplified, turbulent magnetic field, and that energetic
charges at \teq{\gamma \sim \gammax} will possess diffusive mean free
paths \teq{\lambda_{\parallel}\gtrsim 10^6r_g}, an inference concordant
in character with that for our blazar studies here.

%\newpage

\section{Conclusions}

This paper has offered a deep exploration of multi-wavelength spectral
fits to flaring emission from the blazars Mrk 501, BL Lacertae and the
Bl Lac object AO 0235+164.  The modeling uses complete thermal plus
non-thermal electron distribution functions obtained directly from Monte
Carlo simulations of diffusive acceleration at relativistic shocks that
span over six decades in energy, a material advantage over plasma codes.
These simulations presume a planar shock interface and
phenomenologically describe the diffusive mean free path of charges'
interactions with MHD turbulence in a two-dimensional (tensor)
construction, omitting contributions to particle transport that might
emerge from complicated magnetic structures in the shock layer, for
example rippling of the interface.  The importance of treating the
intimate connection between non-thermal and thermal leptons in the jet
is highlighted by the successful modeling of the ephemeral X-ray
component in AO0235+164 as a bulk Comptonization signature of the
thermal population. Our integrated application of simulated relativistic
electron populations to blazar spectral modeling provides fundamentally
new insights into the plasma environment of extragalactic jets.  In
particular, the multiwavelength model fits specify the values of the
magnetic field, plasma density, plasma frequency and Alfv\'en speed
concomitantly.  Restriction to one-zone constructions is made in order
to reduce the number of model parameters. We believe that any extension
of this protocol to encapsulate multiple radiation zones, perhaps
desirable for treating the BL Lac gamma-ray signal, will not alter the
principal inferences we forge concerning the blazar jet environment.

The main conclusions are as follows.  In order to model both the low
synchrotron turnover energy in the optical and X-ray bands, and the
prominent hard gamma-ray flux seen by {\it Fermi} and Atmospheric
\v{C}erenkov Telescopes at even higher energies, leptonic models need
the diffusive mean free path to increase rapidly with electron momentum,
\teq{\lambda_{\parallel}\propto p^{\alpha}} with \teq{\alpha \gtrsim
3/2}, starting with fairly modest values \teq{\lambda_{\parallel}/r_g
\sim 20-100} at low, quasi-thermal momenta.   This establishes mean free
paths orders of magnitude larger than electronic gyroradii at the
Lorentz factors that generate the synchrotron and SSC \teq{\nu F_{\nu}}
peak emission. This determination is robust in that dividing the jet
region into several zones will not ameliorate this constraint on the
\teq{\lambda_{\parallel}(p)} function, since it is principally dictated
by the synchrotron component.  We argue here that this strong momentum
dependence of \teq{\lambda_{\parallel}/r_g} is not unexpected, due to
somewhat constrained inertial ranges of plasma turbulence, and
associated non-gyroresonant diffusion.  It is perhaps also due to a
decline in the level of such turbulence away from shocks that inject
energetic particles into the much larger blazar jet emission zones. 
Such a spatial non-uniformity in jet plasma turbulence spanning small
injection sites to larger dissipative zones in some ways parallels the
much more intimately-observed environment of the active solar wind.
Future investigative programs naturally indicated by this study include
detailed assessment of the particle diffusion characteristics, and their
variation with particle momenta, in relativistic MHD turbulence.

\newpage

\section*{Acknowledgments}

The authors thank Ana Pichel for providing a pre-publication release of 
VERITAS spectral data for the May 2009 flare episode of Mrk 501 that 
served as a guide for our multi-wavelength studies.  We thank Alan Marscher and Greg Madejski 
for a number of helpful comments and careful reading of the manuscript, and the anonymous
referee for constructive suggestions that polished the presentation.  
MGB and MB are grateful to NASA for partial support for this research 
through the Astrophysics Theory Program, grant NNX10AC79G.
MGB also acknowledges support from the Department of
Energy under grant DE-SC0001481 during the early stages of this program.
This work is based on research supported by the South African Research Chairs
Initiative (grant no. 64789) of the Department of Science and Technology and the
National Research Foundation\footnote{Any opinion, finding and 
conclusion or recommendation expressed in this material is that of the authors
and the NRF does not accept any liability in this regard.} of South Africa.

\end{document}